\documentclass[12pt]{article}

\usepackage{amsthm,color}
\usepackage{amsfonts}
\usepackage{graphicx}
\usepackage{mathrsfs}
\usepackage{amsmath}
\usepackage{amssymb}
\usepackage{float}
\usepackage{natbib}
\usepackage{booktabs}
\usepackage{url}
\usepackage{multirow}
\usepackage{arydshln}
\usepackage{courier}
\usepackage{authblk} 

\usepackage{multicol}
\usepackage{latexsym}
\usepackage{psfrag}
\usepackage[usenames,dvipsnames]{xcolor}

\usepackage{breakcites}
\usepackage{bbm}
\usepackage{epsfig,epstopdf}
\usepackage[colorlinks, linkcolor=black, citecolor=black]{hyperref}

\usepackage[utf8]{inputenc}
\usepackage[english]{babel}
\usepackage{caption}
\usepackage{setspace}

\usepackage{sectsty}
\sectionfont{\fontsize{12}{15}\selectfont}
\subsectionfont{\fontsize{12}{15}\selectfont}

\newcommand{\be}{\begin{equation}}
\newcommand{\ee}{\end{equation}}
\newcommand{\ba}{\begin{eqnarray}}
\newcommand{\ea}{\end{eqnarray}}
\newcommand{\bas}{\begin{eqnarray*}}
\newcommand{\eas}{\end{eqnarray*}}
\newtheorem{theorem}{Theorem}
\newtheorem{corollary}{Corollary}
\newtheorem{lemma}{Lemma}

\newtheorem{example}{Example}

\def\proof {{\noindent\it Proof.}\quad}
\def\no{\noindent}

\newcommand{\bsu}{\mbox{\scriptsize \boldmath $u$}}

\newcommand{\var}{{\mbox{Var}}}
\newcommand{\cov}{{\mbox{Cov}}}

\newcommand{\bm}{\mbox{\boldmath $m$}}

\newcommand{\bbeta}{\mbox{\boldmath $\beta$}}

\newcommand{\btheta}{\mbox{\boldmath $\theta$}}

\newcommand{\bpsi}{\mbox{\boldmath $\psi$}}

\newcommand{\boeta}{\mbox{\boldmath $\eta$}}

\newcommand{\bstheta}{\mbox{\scriptsize \boldmath $\theta$}}

\newcommand{\bnu}{\mbox{\boldmath $\nu$}}
\newcommand{\bsnu}{\mbox{\scriptsize \boldmath $\nu$}}

\newcommand{\bGamma}{\mbox{\boldmath $\Gamma$}}

\newcommand{\bLambda}{\mbox{\boldmath $\Lambda$}}

\newcommand{\diag}{\mbox{diag}}
\newcommand{\bq}{\boldsymbol q}
\newcommand{\bQ}{\boldsymbol Q}

\newcommand{\bA}{\boldsymbol A}
\newcommand{\bg}{\boldsymbol g}
\newcommand{\bB}{\boldsymbol B}
\newcommand{\bC}{\boldsymbol C}
\newcommand{\bD}{\boldsymbol D}

\newcommand{\bu}{\boldsymbol u}

\newcommand{\bW}{\boldsymbol W}

\newcommand{\bS}{\boldsymbol S}

\newcommand{\0}{\mbox{\bf 0}}

\newcommand{\bfe}{\boldsymbol e}

\newcommand{\Sum}{\sum_{i=0}^1\sum_{j=1}^{n_{i1}}}

\allowdisplaybreaks

\begin{document}

{\centering {\large {\bf Semiparametric inference on general functionals of two semicontinuous  populations}}\par}
\bigskip

\centerline{Meng Yuan\footnote{Department of Statistics and Actuarial Science, University of Waterloo, Waterloo, Ontario, Canada N2L 3G1.
{\em m33yuan@uwaterloo.ca}},
 \ Chunlin Wang \footnote{Department of Statistics, School of Economics and Wang Yanan Institute for Studies in Economics, Xiamen University, Xiamen, 361005, China. {\em wangc@xmu.edu.cn}}, 
 \ Boxi Lin\footnote{Dalla Lana School of Public Health, University of Toronto, Toronto, Ontario, Canada M5T 3M7. {\em boxi.lin@mail.utoronto.ca}},
 \ and \ Pengfei Li \footnote{Department of Statistics and Actuarial Science, University of Waterloo, Waterloo, Ontario, Canada N2L 3G1. {\em pengfei.li@uwaterloo.ca}}}
 
\bigskip

\bigskip

\hrule

{\small
\begin{quotation}
\no
In this paper, we propose new semiparametric  procedures for making inference on linear functionals and their functions of
two semicontinuous populations.
The distribution of each population is usually characterized by a mixture of a discrete point mass at zero and a continuous skewed positive component, and hence such distribution is semicontinuous in the nature.
To utilize the information from both populations,
we model the positive components of the two mixture distributions via a semiparametric density ratio model.
Under this model setup, we construct the maximum empirical likelihood estimators of the linear functionals and their functions,
and establish the asymptotic normality of the proposed estimators.
We show the proposed estimators of the linear functionals are more efficient than the fully nonparametric ones.
The developed asymptotic results enable us to construct confidence regions and perform hypothesis tests for the linear functionals and their functions. We further apply these results to several important summary quantities such as the moments, the mean ratio, the coefficient of variation, and the generalized entropy class of inequality measures.
Simulation studies demonstrate the advantages of our proposed semiparametric method over some existing methods.
Two real data examples are provided for illustration.

\vspace{0.3cm}

\no
\end{quotation}
}

\hrule

\bigskip

\bigskip

\section{Introduction}
\label{intro}
In this paper, we propose  new statistical inference procedures for linear functionals and their functions of two semicontinuous populations.
Specifically, suppose that we have two independent samples of interest, which are generated by the following mixture models:
\begin{equation}\label{eq:1}
  X_{i1},\cdots, X_{in_i} \sim F_i(x)= v_iI(x \geq 0)+(1-v_i)I(x>0)G_i(x), \quad \mbox{for~}~i=0,1,
\end{equation}
where $v_i\in (0,1)$, $n_i$ is the sample size for the $i$th sample, $I(\cdot)$ is an indicator function, and $G_i(\cdot)$'s are cumulative distribution functions (CDFs) of positive observations in the $i$th sample.
We are interested in estimating linear functionals \citep[p. 6]{fernholz1983mises} of $F_0(x)$ and $F_1(x)$, defined as
\begin{equation}
\label{functional}
\bpsi_0= \int_0^{\infty} {\bf a}(x) dF_0(x)
\quad \mbox{ and } \quad
\bpsi_1= \int_0^{\infty} {\bf a}(x) dF_1(x)
\end{equation}
for some given function ${\bf a}(x)$,
and also functions of $\bpsi_0$ and $\bpsi_1$.
The parameters $\bpsi_0$, $\bpsi_1$, and their functions
include many important summary quantities such as the centered and uncentered moments, the coefficient of variation, the generalized entropy class of inequality measures of each population, and the mean ratio of two such populations as special cases. More details can be found in Section \ref{examples}.
Without loss of generality, we assume that ${\bf a}(0)={\bf 0}$ throughout the paper; this assumption is satisfied by all the examples considered in Section \ref{examples}.

%

Many statistical applications naturally produce semicontinuous data with a mixture of excessive zero values and skewed positive outcomes.
Examples include the medical cost data in public health research \citep{Zhou2000}, and seasonal activity pattern data of field mice  in biological science \citep{mice}.
More examples can be found in \cite{Wang2017},  a special issue of the {\it Biometrical Journal} \citep{Bohning2016},  and references therein.
The parameters $\bpsi_0$, $\bpsi_1$, and their functions  are widely used in many areas.
For example, mean ratio of two populations is a desirable summary {quantity} which characterizes the differences of medical cost in two groups in public health discipline \citep{Zhou2000}.
The moments and  the generalized entropy class of inequality measures
are important summary measures in business and  economic studies \citep{Dufour2019}.

Most existing procedures  for making inference on $\bpsi_0$, $\bpsi_1$, and their functions
are either fully parametric or fully nonparametric.
The parametric procedures are developed under  the parametric model assumption, for example, log-normal assumption,  on $G_i$, $i= 0, 1$.
Under this assumption,
\cite{Tu1999}  and  \cite{ZhouTu1999} developed a Wald-type test and a
likelihood ratio test for the equality of two population means.
Under the same assumption,
\cite{Zhou2000} proposed a maximum likelihood  method and a two-stage bootstrap method
to construct the confidence intervals (CIs) for the mean ratio;
\cite{Chen2006} developed a set of approaches for constructing CIs for the mean ratio based on
the generalized pivotal and likelihood ratio test statistic.
The fully nonparametric methods usually first estimate $F_0(x)$  and $F_1(x)$ by the corresponding empirical CDFs,
which are then used to construct the estimators for $\bpsi_0$, $\bpsi_1$, and their functions.
The asymptotic results of this type of estimators have been well studied in the literature.
See \cite{Serfling1980} for more details.
Nonparametric Wald-type method \citep{Brunner1997,Pauly2015, Dufour2019}
and empirical likelihood (EL) method \citep{kang2010empirical,Wu2012,satter2020jackknife}
may also be used to construct the CIs and perform hypothesis testing for  $\bpsi_0$, $\bpsi_1$, and their functions.


In general, the methods based on the parametric assumption on $G_i$'s are quite efficient.
However, in many applications, the parametric assumption, for example, the log-normal assumption for $G_i$ may be violated.
The corresponding parametric inference results may not be robust to the model misspecification on $G_i$'s \citep{Nixon2004}.
The fully nonparametric methods are generally quite robust to the model assumption on $G_i$'s.
In two-sample setting, the two populations may share certain similar characteristics.
For example, the strength of lumber produced in Canada for different years may follow similar distributions \citep{ChenLiu2013,Cai2017,Cai2018}.
Certain relationship exists between distributions of biomarkers for diagnosing Duchenee Muscular Dystrophy in case and control groups \citep{yuan2020semiparametric}.
The fully nonparametric methods, however,  ignore such information.

In this paper, we propose new semiparametric procedures for estimating $\bpsi_0$, $\bpsi_1$, and their functions
based on the semiparametric density ratio model (DRM) \citep{Anderson1979,Qin2017},
which effectively utilize the information in both populations.
 Let $dG_i(x)$ be  the probability density function of $G_i(x)$, $i=0,1$.
The DRM links
the two CDFs $G_0(x)$ and $G_1(x)$ in model (\ref{eq:1})
as
\begin{equation}
\label{drm}
dG_1(x) = \exp\{\alpha +\boldsymbol{\beta}^\top\bq(x)\}dG_0(x)
\end{equation}
for a pre-specified, non-trivial, basis function $\bq(x)$ of dimension $d$, and unknown parameters $\alpha$ and $\boldsymbol\beta$.
In (\ref{drm}), the baseline distribution $G_0(x)$ is not specified.
Hence the DRM is a  semiparametric model and has the advantage to avoid making risky parametric assumptions on $G_0(x)$ and $G_1(x)$.
 The DRM is also quite flexible and includes many important statistical models as special cases.
For example,
when $\boldsymbol{q}(x)=\log(x)$,  the DRM embraces the log-normal distribution of same variance with respect to the log-scale,
as well as the gamma distribution with the same scale parameters \citep{kay1987transformations}.
{\cite{Jiang2012} pointed out that the DRM is actually broader than Cox proportional hazard models.}
The DRM is also closely related with the well-studied logistic regression \citep{Qin1997}.
The inference under the DRM can be converted to that under the logistic regression \citep{Wang2017}.

The DRM has been proved to be a useful tool for making inference when there is an excess of zeros in the data. \cite{Wang2017,Wang2018} developed the EL ratio (ELR) statistics for testing the homogeneity of distributions and the equality of population means, respectively.
{Under the same setup, 
\cite{lu2020} considered a test for the equality of the zero proportions and the equality of the means of two positive components jointly.}
Their simulation results show that the proposed tests have great power advantages over the existing nonparametric tests.
To our best knowledge, the semiparametric inference procedures such as the point estimation and confidence regions
for the general parameters $\bpsi_0$, $\bpsi_1$, and their functions have not been explored
under the mixture model \eqref{eq:1} and the DRM \eqref{drm}. This paper aims to fill in the void.

Under the mixture model \eqref{eq:1} and the DRM \eqref{drm},
we consider a class of general parameters $\bpsi$ of length $p$,
defined as
\begin{equation}
    \label{psi}
\bpsi= \int_0^\infty \bu(x;\bnu,\btheta) dG_0(x),
\end{equation}
where $\bnu=(\nu_0,\nu_1)^\top$, $\btheta = (\alpha,\bbeta^{\top})^{\top}$, and  $\bu(x;\bnu,\btheta)=\left(u_1(x;\bnu,\btheta),\ldots,u_p(x;\bnu,\btheta)\right)^\top$ is some given $p\times 1$ dimensional function,
and  the parameters defined through ${\bf g}(\bpsi)$, where ${\bf g}(\cdot): p \to q$ is a smooth function of $\bpsi$.
Note that $\bpsi$ covers $\bpsi_0$ and $\bpsi_1$, defined in (\ref{functional}), as special cases.
To see this,
let
\begin{equation}
\label{ufun01}
    \bu(X;\bnu,\btheta)=\left(\begin{array}{c}
       \bu_0(X;\bnu,\btheta)  \\
       \bu_1(X;\bnu,\btheta)
    \end{array} \right)= \left(\begin{array}{c}
(1-\nu_0) {\bf a}(x)  \\
(1-\nu_1) {\bf a}(x) \exp\{\alpha+\bbeta^\top\bq(x)\}
    \end{array} \right).
\end{equation}
Then $\bpsi=\left(\bpsi_0^\top,\bpsi_1^\top\right)^\top$ under the assumption that ${\bf a}(0)={\bf 0}$, as assumed after (\ref{functional}).
The parameters $\bpsi$ and ${\bf g}(\bpsi)$ together cover many important summary quantities.  See Section 2.4 for examples.
Using the EL of \cite{owen},
we construct the maximum EL estimators (MELEs) of $\bpsi$ and ${\bf g}(\bpsi)$.
We establish the asymptotic normality of the MELEs of $\bpsi$ and $\bg(\bpsi)$.
These results enable us to construct confidence regions for $\bpsi$ and $\bg(\bpsi)$ and perform hypothesis testing on $\bpsi$ and $\bg(\bpsi)$.
We apply the results for general $\bpsi$ to $\bpsi_0$ and $\bpsi_1$, and
show that the asymptotic variances of the MELEs of $\bpsi_0$ and $\bpsi_1$
are smaller than or equal to  those of nonparametric estimators of $\bpsi_0$ and $\bpsi_1$.

%
%

The rest of this paper is organized as follows. In Section \ref{estimation}, we first present the MELE of $(\bnu,\btheta)$,
and then propose the MELEs of  $\bpsi$ and  ${\bf g}(\bpsi)$.
We study the asymptotic property of the MELEs of $(\bnu,\btheta)$ as well as the MELEs of  $\bpsi$ and $\bg(\bpsi)$.
These results are applied to the MELEs of $\bpsi_0$ and $\bpsi_1$.
We further provide examples for $\bpsi$ and ${\bf g}(\bpsi)$ which cover several important summary quantities.
Simulation results are presented in Section \ref{simu}   and two real data applications  are given in Section \ref{realdata}.
We conclude the paper with some discussion in Section \ref{conclude}.
For the convenience of presentation, all the technical details are provided in the supplementary material.

\section{Main results}
\label{estimation}

We denote $n_{i0}$ and $n_{i1}$ as the (random) number of zero observations and positive observations, respectively, in each sample $i=0,1$.  Clearly $n_i = n_{i0}+n_{i1}$, for $i=0,1$.
Without loss of generality, we assume that the first $n_{i1}$ observations in group $i$, $X_{i1},\cdots, X_{in_{i1}}$, are positive,
and the rest $n_{i0}$ observations are 0.
We use $n$ to denote the total (fixed) sample size, i.e., $n=n_0+n_1$.

\subsection{Point estimation of $\bpsi$ and ${\bf g}(\bpsi)$} 

We first discuss the maximum EL procedure for estimating the unknown parameters and functions in models  (\ref{eq:1}) and (\ref{drm}).

With the two samples of observations from model (\ref{eq:1}), the full likelihood function is given as
\begin{eqnarray*}
\mathcal{L}_n& =& \prod_{i=0}^1 \left\{ v_i^{n_{i0}}\left(1-v_i\right)^{n_{i1}} \prod_{j=1}^{n_{i1}}dG_i\left(X_{ij}\right)\right\}.
\end{eqnarray*}
Following the EL principle \citep{owen}, we model the  baseline distribution $G_0(x)$ as
\begin{equation} \label{elr.GA}
G_0(x)= \sum_{i=0}^1\sum_{j = 1}^{n_{i1}}p_{ij}I(X_{ij}\le x),
\end{equation}
where $p_{ij} = dG_0(X_{ij})$ for $i=0,1$ and $j=1,\ldots,n_{i1}$.
With (\ref{elr.GA}) and under the DRM (\ref{drm}), the full likelihood function can be rewritten as
\begin{eqnarray*}
\mathcal{L}_n& =& \prod_{i=0}^1  v_i^{n_{i0}}\left(1-v_i\right)^{n_{i1}}\cdot
\left\{\prod_{i=0}^1\prod_{j=1}^{n_{i1}}p_{ij}\right\}
\cdot
\left[ \prod_{j=1}^{n_{11}} \exp\left\{\alpha+\boldsymbol{\beta}^\top\bq(X_{1j})\right\}\right],
\end{eqnarray*}
where $p_{ij}$'s satisfy  the following constraints
\begin{equation}
\label{constraint1}
p_{ij}>0, \quad  \Sum p_{ij}=1, \quad \mbox{and} \quad \Sum p_{ij}\exp\left\{\alpha +\boldsymbol \beta ^\top\bq(X_{ij})\right\}=1.
\end{equation}
These constraints ensure that $G_0(x)$ and $G_1(x)$ are CDFs.

Let $\boldsymbol{P}=\{p_{ij}\}$.
The MELE of $(\boldsymbol{\nu},\boldsymbol{\theta}, \boldsymbol{P})$ is then defined as
$$
(\hat{\bnu},\hat{\boldsymbol{\theta}}, \hat {\boldsymbol{P}})=
\arg\max_{\bsnu, \bstheta, {\boldsymbol{P}} }  \mathcal{L}_n
$$
subject to the constraints in (\ref{constraint1}).
To numerically calculate the MELE, we write the logarithm of the EL function $\mathcal{L}_n$ as
\begin{equation}\label{lik}
\tilde \ell(\bnu,\btheta, G_0) = \ell_0\left(\bnu\right) +  \tilde \ell_1\left(\boldsymbol{\theta},\boldsymbol{P}\right),
\end{equation}
where
\begin{equation*}
    \ell_0\left(\bnu\right) = \sum_{i=0}^1\log\left\{ v_i^{n_{i0}}\left( 1-v_i\right)^{n_{i1}} \right\}
    ~\text{and}~
    \tilde \ell_1\left(\boldsymbol \theta,\boldsymbol{P}\right)  =  \sum_{j=1}^{n_{11}}\left\{\alpha +\boldsymbol \beta ^\top\bq(X_{1j})\right\}+\sum_{i=0}^1\sum_{j=1}^{n_{i1}}\log p_{ij}.
\end{equation*}
Here $\ell_0\left(\bnu\right)$ is the binomial log-likelihood function corresponding to the number of zero observations and $\tilde \ell_1\left(\boldsymbol \theta, \boldsymbol{P}\right)$ represents the empirical log-likelihood function associated with the positive observations.

Following \cite{Wang2017}, we have $\hat{\bnu} = \arg\max_{\bsnu} \ell_0(\bnu)$ and
\begin{equation*}
(\hat \btheta,\hat {\boldsymbol{P}} ) = \arg\max_{\bstheta,{\boldsymbol{P}} }\left\{\tilde \ell_1\left(\boldsymbol \theta,\boldsymbol{P}\right): ~
p_{ij}>0, ~\Sum p_{ij}=1,~\Sum p_{ij}\exp\left\{\alpha +\boldsymbol \beta ^\top\bq(X_{ij})\right\}=1\right\}.
\end{equation*}
By the method of Lagrange multiplier, $\hat \btheta$ can be obtained by maximizing the following dual empirical log-likelihood function \citep{Cai2017}:
\begin{equation*}
\ell_1(\btheta)=
- \Sum
  \log\left\{ 1 + \hat{\rho}[\exp\{\alpha + \bbeta^{\top} \bq(X_{ij})\}-1] \right\}
+ \sum_{j=1}^{n_{11}} \{\alpha + \bbeta^{\top}  \bq(X_{1j}) \},
\end{equation*}
where $\hat{\rho} = n_{11}\{n_{01}+n_{11}\}^{-1}$.
That is,  $\hat{\btheta}= \arg \max_{\bstheta}\ell_1(\btheta)$.
Note that $\hat\rho$ is a random variable under our setup.
This is fundamentally different from the case when there is no excess of zeros in the data \citep{Qin1997}, and it creates new theoretical challenges for our asymptotic development in the next section.

Once $\hat {\boldsymbol{\theta}} $ is obtained,
the MELEs of $\hat p_{ij}$'s are given as \citep{Wang2017}
$$
\hat{p}_{ij}= \{n_{01}+n_{11}\}^{-1}\left\{1 + \hat{\rho}[\exp\{\hat\alpha + \hat\bbeta^{\top} \bq(X_{ij})\}-1] \right\}^{-1},
$$
and
the MELEs of $G_0(x)$ and $G_1(x)$ are
\begin{equation*}
    \hat G_0(x)=\Sum \hat{p}_{ij} I(X_{ij}\leq x)~~~\text{and}~~~\hat G_1(x)=\Sum \hat{p}_{ij}\exp\{\hat{\alpha} + \hat\bbeta^{\top} \bq(X_{ij})\} I(X_{ij}\leq x).
\end{equation*}

By the definition of $\bpsi$ in \eqref{psi},  $\bpsi$ is a function of $(\bnu,\btheta)$ and $G_0$.
With MELEs $(\hat \bnu,\hat \btheta)$ and $\hat G_0$, the MELE of $\bpsi$ is given by
\ba
\label{hat.psi}
    \hat{\bpsi} =
    \sum_{i=0}^1\sum_{j=1}^{n_{i1}} \hat p_{ij}\bu(X_{ij};\hat\bnu,\hat\btheta),
\ea
and the estimator of $\bg(\bpsi)$ is $\bg(\hat\bpsi)$.

{When $\bu(x;\bnu, \btheta)$ takes the specific form as in (\ref{ufun01}), we obtain the MELEs of $\bpsi_0$ and $\bpsi_1$, defined in (\ref{functional})}, as
\ba
\label{hat.psi01}
 \hat{\bpsi}_0 &=& \sum_{i=0}^1\sum_{j=1}^{n_{i1}} \hat p_{ij}(1-\hat \nu_0) {\bf a}(X_{ij}) ~~\mbox{ and }~~
 \hat{\bpsi}_1 = \sum_{i=0}^1\sum_{j=1}^{n_{i1}} \hat p_{ij}(1-\hat \nu_1) {\bf a}(X_{ij})\exp\{\hat\alpha+\hat\bbeta^\top\bq(x)\}.
\ea


\subsection{Asymptotic properties}
In this section, we first study the asymptotic properties of $\hat\boeta=(\hat\bnu^\top,\hat\rho,\hat\btheta^\top)^\top$ and then apply these results
to establish the asymptotic properties of $\hat{\bpsi}$ and $\bg(\hat\bpsi)$.

%

For ease of presentation, we introduce some notations.
We use $\bnu^{*}$ and $\btheta^{*}$ to denote the true values of $\bnu$ and $\btheta$, respectively.
Let $\bQ(x) = (1,\bq(x)^\top)^\top$ and
\bas
&w = n_0/n,~\Delta^*=w(1-\nu_0^*)+(1-w)(1-\nu_1^*),
~\rho^*=\frac{(1-w)(1-\nu_1^*)}{\Delta^*},\\
&\omega(x)=\exp\{\btheta^{*\top} \bQ(x)\},~h(x)=1+\rho^* \{\omega(x)-1\},~h_1(x)=\rho^* \omega(x)/h(x),\\
& \bA_{\bsnu} =
\diag\left\{ \frac{w}{\nu_0^*(1-\nu_0^*)}, \frac{1-w}{\nu_1^*(1-\nu_1^*)} \right\},~
\bA_{\bstheta} =
\Delta^*(1-\rho^*)E_0\left\{h_1(X) \bQ(X)\bQ(X)^\top \right\},
\eas
where $E_0(\cdot)$ represents the expectation operator with respect to $G_0$ and $X$ refers to a random variable from $G_0$. 
Note that although $\omega(\cdot)$, $h(\cdot)$, and $h_1(\cdot)$ also depend on  $\btheta^*$ and/or $\rho^*$, we drop these redundant parameters for notational simplicity.

The asymptotic results in this section are developed under the following regularity conditions.
\begin{itemize}
    \item [C1:] The true value $\bnu_i^*$ lies between 0 and 1 for $i  = 0,1$.
    \item [C2:] As the total sample size $n$ goes to infinity, $n_0/n = w$  for some constant $w \in (0,1)$.
    \item [C3:] The components of $\bQ(x)$ are continuous and stochastically
linearly independent.
    \item [C4:] $\int_0^{\infty} \exp\{\bbeta^\top\bq(x)\}dG_0(x) < \infty$ for all $\bbeta$ in a neighbourhood of the true value $\bbeta^*$.
\end{itemize}

Condition C1 ensures the binomial likelihood $\ell_0(\bnu)$ has regular properties.
Condition C2 means that both $n_0$ and $n_1$ go to $\infty$ at the same rate.
Conditions C1 and C2 imply that $\bA_{\bsnu}$ is positive definite.
Condition C3 ensures that no linear combinations of any components of $\bQ(x)$ can be 0 with probability 1 under $G_0$.
Condition C4 guarantees the existence of finite moments of $\bq(X)$ in a neighborhood of $\bbeta^*$ under both $G_0(x)$ and $G_1(x)$.
Conditions C3 and C4 together imply that $\bA_{\bstheta}$ is positive definite.

The following theorem establishes the asymptotic normality of $\hat{\boeta}$.
\begin{theorem}
\label{thm1}
Let  $\boeta^*=(\bnu^{*\top},\rho^*,\btheta^{*\top})^\top$.
Assume that Conditions C1--C4 are satisfied.
As the total sample size $n\to\infty$,
\bas
n^{1/2}(\hat\boeta-\boeta^*)
\to
N\left(\0, \bLambda \right)
\eas
in distribution, where
\begin{equation*}
\bLambda=
\left( \begin{array}{ccc}
\bA_{\bsnu}^{-1} & \rho^*(1-\rho^*) \bA_{\bsnu}^{-1}\bW^\top & \0 \\
\rho^*(1-\rho^*) \bW \bA_{\bsnu}^{-1} & (\Delta^*)^{-1}\rho^*(1-\rho^*) \{\rho^*\nu_0^*+(1-\rho^*)\nu_1^*\} & \0 \\
\0 & \0 & \bA_{\bstheta}^{-1}-\frac{\bfe\bfe^\top}{\Delta^{*}\rho^*(1-\rho^*)} \\
\end{array}\right)
\end{equation*}
with $\bW = \left( (1-\nu_0^*)^{-1}, -(1-\nu_1^*)^{-1} \right)$ and $\bfe=(1,\0_{d\times1}^\top)^\top$.
\end{theorem}

\cite{Qin1997} considered the asymptotic normality of $\sqrt{n}(\hat\btheta-\btheta^*)$ 
when there is no excess of zeros in the data.
Theorem \ref{thm1} generalizes their results to the case when the data contains excessive zeros.
Furthermore, it establishes the joint limiting distribution of $\sqrt{n}(\hat\btheta-\btheta^*)$,  $\sqrt{n}(\hat\bnu-\bnu^*)$, and $\sqrt{n}(\hat\rho-\rho^*)$,
where the latter two are induced by the semicontinuous data structure.
This joint limiting distribution plays an important role in deriving the asymptotic normality of $\hat\bpsi$  in  the following theorem.

%
\begin{theorem}
\label{thm2}
Let $\bpsi^*$ be the true value of $\bpsi$. Under the same conditions of Theorem \ref{thm1},
 as $n\to\infty$,
(a)
$
\sqrt{n}(\hat\bpsi - \bpsi^*) \to N(\0, \bGamma)
$
in distribution, where
\begin{eqnarray*}
\label{bgamma.form}
\bGamma= \frac{1}{\Delta^*}E_0\left\{\frac{\bu(X;\bnu^*,\btheta^*)\bu(X;\bnu^*,\btheta^*)^\top}{h(X)}\right\}
- \frac{\bpsi^*\bpsi^{*\top}}{\Delta^*}
+ \mathcal{M}_1  \bA_{\bsnu}^{-1} \mathcal{M}_1 ^\top
- \frac{\mathcal{M}_2  \mathcal{M}_2^\top}{\Delta^*\rho^*(1-\rho^*)}
+ \mathcal{M}_3  \bA_{\bstheta}^{-1} \mathcal{M}_3 ^\top,
\end{eqnarray*}
with
\bas
\label{CM1}
\mathcal{M}_1&=& E_0\left\{ \frac{\partial \bu(X;\bnu^*,\btheta^*)}{\partial\bnu} \right\}, \\
\label{CM2}\mathcal{M}_2&=& E_0\left[\left\{\partial\bu(X;\bnu^*,\btheta^*)/\partial\btheta\right\}\bfe\right]- \rho^*\bpsi^*,\\
\label{CM3}\mathcal{M}_3&=&   E_0\left\{\partial\bu(X;\bnu^*,\btheta^*)/\partial\btheta -
h_1(X)\bu(X;\bnu^*,\btheta^*)\bQ(X)^\top\right\};
\eas
(b)  for some smooth function ${\bf g}(\cdot): p \to q$,
$
\sqrt{n}\left\{ \bg(\hat\bpsi) - \bg(\bpsi^*)\right\} \to N\left(0, \bGamma_{\bg} \right)
$
in distribution, where
$$
\bGamma_{\bg} = \left\{\frac{\partial \bg(\bpsi^*)}{\partial \bpsi}\right\}   \bGamma \left\{\frac{\partial \bg(\bpsi^*)}{\partial \bpsi}\right\}^\top.
$$
\end{theorem}

\cite{li2018comparison} derived a similar result in their Theorem 2.1 for $\hat\bpsi$ when there is no excess of zeros in the data and $p=1$.
Theorem \ref{thm2} covers the case when there exists excessive zeros.
The two results complement each other to cover both cases.

We now apply the results for $\hat\bpsi$ in Theorem \ref{thm2} to $\hat\bpsi_0$ and $\hat \bpsi_1$ in (\ref{hat.psi01}),
and compare them with the fully nonparametric estimators $\tilde\bpsi_0$  and $\tilde \bpsi_1$:
 $$
 \tilde\bpsi_0=\frac{1}{n_0}\sum_{j=1}^{n_0} {\bf a}(X_{0j})
\quad \mbox{ and }\quad
 \tilde\bpsi_1=\frac{1}{n_1}\sum_{j=1}^{n_1} {\bf a}(X_{1j}).
 $$
For $i=0,1$, let
 $$
 {\bf V}_i=\int_{0}^\infty {\bf a}(x)\{ {\bf a}(x) \}^\top dF_i(x)- \int_{0}^\infty {\bf a}(x)dF_i(x) \int_{0}^\infty \{ {\bf a}(x) \}^\top dF_i(x) .
 $$
Then
 $
 \sqrt{n}\left(  \tilde\bpsi_0^\top-\bpsi_0^\top, \tilde\bpsi_1^\top-\bpsi_1^\top \right)^\top
 $
has the asymptotic variance-covariance matrix
 $$
  \bGamma_{non}=\left(
  \begin{array}{cc}
  w^{-1}{\bf V}_0  & \0 \\
         \0 &  (1-w)^{-1} {\bf V}_1
    \end{array}
  \right).
 $$
In comparison with the asymptotic variance of the MELEs $\hat\bpsi_0$ and $\hat \bpsi_1$ given in (\ref{hat.psi01}), we have the following results.
\begin{corollary}
\label{thm3}
Under the same conditions of Theorem \ref{thm1},
 as $n\to\infty$,
\bas
\sqrt{n}
\left(
  \begin{array}{c}
\hat\bpsi_0-\bpsi_0  \\
\hat\bpsi_1-\bpsi_1\\
    \end{array}
  \right)
 \to N(\0, \bGamma_{sem})
\eas
in distribution,
where
$$
\bGamma_{sem}=
  \bGamma_{non}-\Delta^*(1-\rho^*) E_0\left\{h_1(X)
   \left(
  \begin{array}{c}
  w^{-1} {\bf d}(X)\\
  -(1-w)^{-1}{\bf d}(X)\\
  \end{array}
  \right)
  \left(
    \begin{array}{c}
  w^{-1} {\bf d}(X)\\
  -(1-w)^{-1}{\bf d}(X)\\
  \end{array}
  \right)^\top
  \right\},
  $$
with
 $$
{\bf d}(X)={\bf a}(X)- \Delta^*(1-\rho^*) E_0\left\{h_1(X){\bf a}(X)\bQ(X)^\top\right\} \bA_{\bstheta}^{-1} \bQ(X).
$$

\end{corollary}

Corollary \ref{thm3} implies that $  \bGamma_{non}- \bGamma_{sem}$ is positive semidefinite.
Hence the proposed MELEs of $\bpsi_0$ and $\bpsi_1$
are more efficient than the corresponding nonparametric ones.
Simulation studies in Section 3 further confirm this property.


\subsection{Confidence regions and hypothesis tests for $\bpsi$ and  $\bg(\bpsi)$}
The two variance-covariance matrices $\bGamma$ and $\bGamma_{\bg}$
may depend on $\bpsi^*$ and $G_0(x)$.
Replacing them by $\hat\bpsi$ and  $\hat G_0(x)$,
we get the corresponding estimators $\hat\bGamma$ and $\hat\bGamma_{\bg}$.
With the results in Theorem \ref{thm1}, it can be easily shown that both
$\hat\bGamma$ and $\hat\bGamma_{\bg}$ are consistent. The details are hence omitted.

\begin{theorem}
\label{coro1}
Under the same conditions of Theorem \ref{thm1}, as $n \to \infty$,
$\hat\bGamma\to \bGamma$
and
$\hat\bGamma_{\bg}\to \bGamma_{\bg}$
both in probability.
\end{theorem}

Theorems \ref{thm2} and \ref{coro1} together imply that
$$
(\hat \bpsi-\bpsi^*)^\top \hat \bGamma^{-1}(\hat \bpsi-\bpsi^*)
\quad \mbox{ and }\quad
\{ \bg(\hat \bpsi)-\bg(\bpsi^* )\}^\top \hat \bGamma_{\bg}^{-1}\{ \bg(\hat \bpsi)-\bg(\bpsi^* )\}
$$
converge in distribution to $\chi^2_p$ and $\chi^2_q$, respectively.
Hence both of them are asymptotically pivotal and can be used to construct  Wald-type
confidence regions for $\bpsi$ and $\bg(\bpsi)$ and preform hypothesis tests about $\bpsi$ and $\bg(\bpsi)$.
For illustration, we consider the case when the dimension  $q$ of ${\bf g}(\cdot)$ is 1, which is perhaps the most common situation in applications.
 Let $\phi=\bg(\bpsi)$.
Next, we explain how to apply the obtained results to construct a $100(1-\gamma)\%$ CI for $\phi$
and perform the hypothesis test for $H_0: \phi=0$.
For general  $\bpsi$ and $\bg(\bpsi)$, the procedures can be followed similarly.

Let $\hat\phi=\bg(\hat\bpsi)$ and $\hat\sigma_\phi^2=\hat\bGamma_{\bg}$.
Then a $100(1-\gamma)\%$ CI for $\phi$ is
\begin{equation}
\label{CI.phi}
\mathcal{I}_{\phi}=\left\{\phi: (\hat\phi-\phi)^2/\hat\sigma_\phi^2\leq\chi^2_{1,\gamma}  \right\}
=\left[\hat\phi-z_{\gamma/2}\hat\sigma_\phi,\hat\phi+z_{\gamma/2}\hat\sigma_\phi\right],
\end{equation}
where $\chi^2_{1,\gamma}$ and $z_{\gamma/2}$ denote the ($1-\gamma$) quantile of $\chi^2_1$ distribution and the ($1-\gamma/2$) quantile of $N(0,1)$ distribution, respectively.
For testing $H_0: \phi=0$,
we reject the null hypothesis if
\begin{equation}
\label{test.phi}
\hat\phi ^2/\hat\sigma_\phi^2>\chi^2_{1,\gamma}
~~
\mbox{
or equivalently
}
~~
|\hat\phi/\hat\sigma_\phi|>z_{\gamma/2}
\end{equation}
under the given significance level $\gamma$.
%
%


\subsection{Examples of $\bpsi$ and $\bg(\bpsi)$}
\label{examples}

In this section, we provide some examples to demonstrate that $\bpsi$ and $\bg(\bpsi)$ cover many important summary quantities.
The proposed methods and the general results in Sections 2.1--2.3 can be readily applied to these quantities.

\begin{example} (Uncentered moments)
Let $\mu_i^{(k)}=\int_{0}^\infty x^k dF_i(x)$ be the $k$th (uncentered) moments of $F_i(x)$, $i=0,1$.
When $k=1$, we write $\mu_i=\mu_i^{(1)}$.
Clearly, when
\begin{equation}
\label{ufun.moments}
u_1(x;\bnu,\btheta)=(1-\nu_0)x^k~~~\text{and}~~~u_2(x;\bnu,\btheta) = (1-\nu_1)x^k\exp\{\alpha + \bbeta^{\top} \bq(x)\},
\end{equation}
then
$\bpsi=(\mu_0^{(k)},\mu_1^{(k)})^\top$.
\end{example}

\begin{example} (Mean ratio)
\label{example2}
Let $\delta=\mu_1/\mu_0$ denote the mean ratio of two populations.
Setting $k=1$ in (\ref{ufun.moments}), we obtain $\bpsi=(\mu_0,\mu_1)^\top$.
Further let $g(x_1,x_2)=x_2/x_1$, then we get $\delta=g(\bpsi)$.
We can directly construct a CI of $\delta$ by using the result given in (\ref{CI.phi}).

An alternative way is to consider $g(x_1,x_2)=\log(x_2)- \log( x_1)$, then we get $g(\bpsi)=\log \delta$.
We can use the form of (\ref{CI.phi}) to construct a CI for $\log\delta$ first and then transform it to the CI for $\delta$.
Our simulation indicates that the second way leads to a CI with better coverage accuracy.
\end{example}

\begin{example} (Centered moments)
\label{example3}
Let $C_i^{(k)}=\int_{0}^\infty (x-\mu_i) ^k dF_i(x)$ be the $k$th centered moments of $F_i(x)$, $i=0,1$.
When $k=2$, we write $\sigma_i^2=C_i^{(2)}$.
As demonstrated in \cite{Serfling1980}, centered moments $C_i^{(k)}$ can be written as functions of $\mu_i^{(1)},\ldots, \mu_i^{(k)}$.
For illustration, we concentrate on $k=2$ and consider the variances of the two populations, $\sigma_0^2$ and $\sigma_1^2$.

Let
\begin{equation*}
\bu(x;\bnu,\btheta)=\left( (1-\nu_0)x, (1-\nu_0)x^2, (1-\nu_1)x\exp\{\alpha + \bbeta^{\top} \bq(x)\},  (1-\nu_1)x^2\exp\{\alpha + \bbeta^{\top} \bq(x)\}\right)^\top,
\end{equation*}
then
$\bpsi=(\mu_0, \mu_0^{(2)},\mu_1,\mu_1 ^{(2)} )^\top$.
Define $\bg(\cdot)$ as
$$
\bg(x_1,x_2,x_3,x_4)=(x_2-x_1^2,x_4-x_3^2)^\top.
$$
We have $\bg(\bpsi)=(\sigma_0^2,\sigma_1^2)^\top$.
The results in Theorem \ref{thm2} can be used to obtain the joint limiting distribution of $\sqrt{n}(\hat\sigma_0^2-\sigma_0^2, \hat\sigma_1^2-\sigma_1^2)^\top$, where $\hat\sigma_0^2$ and $\hat\sigma_1^2$ are the MELEs of $\sigma_0^2$ and $\sigma_1^2$, respectively.

If we choose
$$
\bg(x_1,x_2,x_3,x_4)=(x_4-x_3^2)-(x_2-x_1^2).
$$
Then $\bg(\bpsi)= \sigma_1^2-\sigma_0^2$, and the procedure described in (\ref{test.phi}) can be used to test $H_0:\sigma_0^2=\sigma_1^2$.

\end{example}

\begin{example} (Coefficient of variation)
\label{example4}
Let $CV_i=\sigma_i/\mu_i$ be the coefficient of variation of the $i$th population, $i=0,1$.
Suppose that the same $\bu(\cdot)$ function specified in Example \ref{example3} is used. If we choose $\bg(\cdot)$ to be
$$
\bg(x_1,x_2,x_3,x_4)=( \sqrt{x_2}/x_1,\sqrt{x_4}/x_3)^\top,
$$
then $\bg(\bpsi)=(CV_0,CV_1)^\top$.

If we choose $\bg(x_1,x_2,x_3,x_4)=\sqrt{x_4}/x_3- \sqrt{x_2}/x_1 $, then $\bg(\bpsi)=CV_1-CV_0$, and  the procedure described in (\ref{test.phi}) can be used to test $H_0:CV_0=CV_1$.

\end{example}

\begin{example} (Generalized entropy class of inequality measures)
Let
$$
GE_{i}^{(\xi)}
=\left\{
\begin{array}{ll}
\frac{1}{\xi^2-\xi}\left\{ \int_{0}^\infty \left( \frac{x}{\mu_i} \right )^\xi d F_i(x)-1 \right\},&\mbox{if }\xi\neq0,1,\\
-\int_{0}^\infty \log\left(\frac{x}{\mu_i}\right )dF_i(x),&\mbox{if }\xi=0,\\
\int_{0}^\infty  \frac{x}{\mu_i} \log\left(\frac{x}{\mu_i}\right )dF_i(x) ,&\mbox{if }\xi=1,\\
\end{array}
\right.
$$
be the generalized entropy class of inequality measures of the $i$th population, $i=0,1$.
Note that $GE_{i}^{(\xi)}$ are not well-defined for the population with excessive zeros when $\xi=0$.
Under our setup,  $(GE_0^{(\xi)},GE_1^{(\xi)})^\top$ can also be written as $\bg(\bpsi)$
with certain $\bu(\cdot)$ and $\bg(\cdot)$ functions as long as $\xi\neq 0$.
For illustration, we consider $\xi=1$.

Let
\begin{equation*}
\bu(x;\bnu,\btheta)=\left( (1-\nu_0)x, (1-\nu_0)x \log(x), (1-\nu_1)x\exp\{\alpha + \bbeta^{\top} \bq(x)\},  (1-\nu_1)x\log(x) \exp\{\alpha + \bbeta^{\top} \bq(x)\}\right)^\top
\end{equation*}
and
$$
\bg(x_1,x_2,x_3,x_4)=( x_2/x_1-\log x_1,x_4/x_3-\log x_3)^\top.
$$
Then  $\bg(\bpsi)=( GE_0^{(1)}, GE_1^{(1)} )^\top$.
Similar to Examples \ref{example3} and \ref{example4}, we can choose an appropriate $\bg(\cdot)$ function to construct a testing procedure
for $H_0: GE_0^{(1)}=GE_1^{(1)}$.
\end{example}

\section{Simulation study}
\label{simu}
In this section, we conduct simulations to compare the finite sample performance of our proposed estimators and CIs with some existing methods.
We consider three parameters, the mean ratio $\delta$, discussed in Example \ref{example2},
and the population variances $\sigma^2_0$ and $\sigma^2_1$, discussed in Example \ref{example3},
for performance comparison of point estimators.
For comparison of CIs, we mainly focus on the mean ratio $\delta$.

\subsection{Simulation setup}

In our simulations, the random observations are generated from the mixture model (\ref{eq:1}), with $G_i$ being the log-normal distribution.
We use log-normal distribution in simulations because it has positive support and is highly skewed to the right.
These properties allow us to check the applicability of the proposed method for skewed data which is commonly seen in reality.
We use $\mathcal{LN}(a , b)$ to denote a log-normal distribution, where $a $ and $b $ are respectively the mean and variance in the log scale. The parameter settings for simulation studies are given in Table \ref{tab:1}.

\begin{table}[!htt]
\caption{Parameter settings for simulation studies: $G_0=\mathcal{LN}(a_0 , b_0)$ and $G_1=\mathcal{LN}(a_1 , b_1)$.}
\tabcolsep 1mm
\scriptsize
\centering
\begin{tabular}{ccccccc}
  \hline
Model & $(v_0,v_1)$ & $(a_0,a_1)$ & $(b_0,b_1)$ & $(\mu_0,\mu_1)$ & $(\sigma_0^2,\sigma_1^2)$ & $\delta$ \\
  \hline
1 & (0.30, 0.30) & (0.00, 0.00) & (1.00, 1.00) & (1.15, 1.15) & (3.84, 3.84) & 1.00 \\
  2 & (0.70, 0.70) & (0.00, 0.00) & (1.00, 1.00) & (0.49, 0.49) & (1.97, 1.97) & 1.00 \\
  3 & (0.30, 0.50) & (0.33, 0.66) & (1.00, 1.00) & (1.61, 1.59) & (7.43, 11.29) & 0.99 \\
  4 & (0.50, 0.70) & (0.37, 0.89) & (1.00, 1.00) & (1.19, 1.20) & (6.32, 11.69) & 1.01 \\
  5 & (0.50, 0.30) & (0.00, 0.00) & (1.00, 1.00) & (0.82, 1.15) & (3.02, 3.84) & 1.40 \\
  6 & (0.70, 0.50) & (0.00, 0.00) & (1.00, 1.00) & (0.49, 0.82) & (1.97, 3.02) & 1.67 \\
  7 & (0.60, 0.40) & (0.00, 0.00) & (1.00, 1.00) & (0.66, 0.99) & (2.52, 3.45) & 1.50 \\
  8 & (0.30, 0.30) & (0.00, 0.50) & (1.00, 1.00) & (1.15, 1.90) & (3.84, 10.44) & 1.65 \\
  9 & (0.70, 0.70) & (0.00, 0.75) & (1.00, 1.00) & (0.49, 1.05) & (1.97, 8.84) & 2.12 \\
  10 & (0.40, 0.60) & (0.00, 1.00) & (1.00, 1.00) & (0.99, 1.79) & (3.45, 18.63) & 1.81 \\
   \hline
\end{tabular}
\label{tab:1}
\end{table}

For all the models considered in Table \ref{tab:1}, the DRM \eqref{drm} is satisfied with $\boldsymbol q(x)=\log x$.
For each model, we consider four combinations of sample sizes $(n_0, n_1)$:  $(50,50)$, $(100,100)$, $(50,150)$, and $(150,50)$.
The number of replications is 10,000 for each configuration of the parameter settings.

\subsection{Comparison for point estimators}
We first study the finite sample performance of point estimators. Under the model \eqref{eq:1} and DRM \eqref{drm}, our proposed estimators for $\delta$, $\sigma_0^2$, and $\sigma_1^2$ are
\begin{equation*}
    \hat{\delta} = \frac{\hat\mu_1}{\hat\mu_0},~~
    \hat \sigma_0^2 = (1-\hat v_0)\Sum \hat p_{ij}X_{ij}^2-\hat \mu_0^2,
    ~~\text{and}~~
    \hat \sigma_1^2 = (1-\hat v_1)\Sum \hat p_{ij}\exp\left\{\hat\alpha+\boldsymbol{\hat\beta}^\top\mathbf{q}(X_{ij})\right\} X_{ij}^2-\hat\mu_1^2,
\end{equation*}
respectively, with
\begin{equation*}
    \hat\mu_0 = (1-\hat v_0)\sum_{i=0}^1\sum_{j=1}^{n_{i1}}\hat p_{ij}X_{ij}
    ~~\text{and}~~
    \hat \mu_1 = (1-\hat v_1)\sum_{i=0}^1\sum_{j=1}^{n_{i1}}\hat p_{ij}\exp\left\{\hat\alpha+\hat\bbeta^\top\bq(X_{ij})\right\}X_{ij}.
\end{equation*}

We use simulation studies to compare the proposed  estimators $\hat\delta$, $\hat\sigma^2_0$, and  $\hat\sigma^2_1$ with the fully nonparametric estimators
\begin{equation*}
  \tilde\delta= \frac{\tilde \mu_1}{\tilde \mu_0}, ~~
  \tilde \sigma^2_i = \frac{1}{n_i-1}\sum_{j = 1}^{n_i}(X_{ij}-\tilde \mu_i)^2
  ~~\text{with}~~
  \tilde \mu_i=\frac{1}{n_i}\sum_{j = 1}^{n_i}X_{ij},
\end{equation*}
for $i  = 0,1$.

The performance of a point estimator is evaluated in terms of the bias and mean square error (MSE).
The simulation results are summarized in Table \ref{tabratio}.

\begin{table}[!htt]
\caption{Bias and mean square error of point estimates for $\delta$, $\sigma^2_0$, and $\sigma^2_1$.}
\centering
\tabcolsep 1mm
\scriptsize
{
\begin{tabular}{ cl | cc cc | cccc | cccc }
  \hline
    &  & \multicolumn{2}{c}{$\tilde \delta$} &  \multicolumn{2}{c |}{$\hat\delta$}& \multicolumn{2}{c}{$\tilde \sigma_0^2$} & \multicolumn{2}{c|}{$\hat\sigma_0^2$} &
         \multicolumn{2}{c}{$\tilde\sigma_1^2$} & \multicolumn{2}{c }{$\hat\sigma_1^2$} \\
 \hline
 Model & $(n_0,n_1)$& Bias  & MSE & Bias  & MSE& Bias  & MSE& Bias  & MSE& Bias  & MSE& Bias  & MSE \\
 \hline
  \multirow{4}{*}{1} &  (50, 50)  & 0.06 & 0.13 & 0.04 & 0.09 &  0.05 &  39.62 & -0.03 & 21.72 &  0.03 &   34.87 & -0.01 &   22.25 \\
   &  (50, 150)  & 0.06 & 0.09 & 0.03 & 0.06 & -0.04 &  23.11 &  0.01 &  8.38 &  0.02 &    9.23 & -0.03 &    7.07 \\
   &  (150, 50)  & 0.01 & 0.08 & 0.02 & 0.05 & -0.01 &  10.74 & -0.08 &  8.09 & -0.02 &   50.72 &  0.06 &   18.71 \\
   &  (100, 100)  & 0.03 & 0.06 & 0.02 & 0.04 & -0.01 &  17.81 & -0.05 & 10.12 & -0.03 &   14.53 & -0.06 &    8.14 \\ \hline
  \multirow{4}{*}{2} &  (50, 50)  & 0.17 & 0.59 & 0.13 & 0.39 &  0.09 &  61.65 &  0.03 & 39.30 & -0.01 &   20.32 & -0.02 &   15.92 \\
   &  (50, 150)  & 0.18 & 0.41 & 0.11 & 0.26 & -0.01 &   9.91 &  0.09 &  6.71 &  0.02 &    6.63 & -0.03 &    4.58 \\
   &  (150, 50)  & 0.06 & 0.25 & 0.06 & 0.18 &  0.00 &   3.62 & -0.04 &  2.86 &  0.04 &   14.89 &  0.10 &    6.11 \\
   &  (100, 100)  & 0.09 & 0.21 & 0.06 & 0.15 & -0.01 &   5.85 & -0.01 &  3.72 &  0.01 &    5.51 & -0.03 &    3.19 \\ \hline
  \multirow{4}{*}{3} &  (50, 50)  & 0.06 & 0.17 & 0.03 & 0.11 & -0.05 & 104.16 & -0.03 & 54.97 &  0.19 &  426.17 & -0.15 &  292.11 \\
   &  (50, 150)  & 0.06 & 0.10 & 0.03 & 0.07 & -0.05 & 118.37 &  0.19 & 36.01 &  0.11 &  120.63 & -0.08 &  100.58 \\
   &  (150, 50)  & 0.01 & 0.11 & 0.02 & 0.08 &  0.02 &  42.75 & -0.12 & 23.86 & -0.26 &  193.29 & -0.16 &  135.72 \\
   &  (100, 100)  & 0.03 & 0.08 & 0.02 & 0.05 & -0.06 &  57.48 & -0.10 & 21.76 & -0.09 &  147.69 & -0.21 &  108.06 \\ \hline
  \multirow{4}{*}{4} &  (50, 50)  & 0.09 & 0.33 & 0.07 & 0.24 &  0.08 & 147.42 &  0.03 & 54.06 & -0.04 &  458.52 & -0.31 &  398.77 \\
   &  (50, 150)  & 0.09 & 0.19 & 0.05 & 0.14 & -0.02 &  72.97 &  0.24 & 26.55 & -0.08 &  160.82 & -0.27 &  138.70 \\
   &  (150, 50)  & 0.03 & 0.21 & 0.04 & 0.16 &  0.00 &  32.66 & -0.09 & 18.79 &  0.02 &  438.73 & -0.02 &  289.92 \\
   &  (100, 100)  & 0.04 & 0.14 & 0.03 & 0.11 &  0.03 &  57.22 &  0.01 & 19.11 &  0.03 &  275.65 & -0.11 &  233.74 \\ \hline
  \multirow{4}{*}{5} &  (50, 50)  & 0.13 & 0.38 & 0.08 & 0.24 &  0.01 &  32.45 & -0.02 & 13.00 & -0.04 &   25.04 & -0.12 &   19.46 \\
   &  (50, 150)  & 0.13 & 0.28 & 0.07 & 0.18 & -0.05 &  16.82 &  0.03 &  6.33 & -0.07 &    8.81 & -0.13 &    6.94 \\
   &  (150, 50)  & 0.04 & 0.18 & 0.04 & 0.12 &  0.00 &   7.68 & -0.05 &  5.76 & -0.04 &   41.11 &  0.02 &   16.64 \\
   &  (100, 100)  & 0.06 & 0.16 & 0.04 & 0.10 &  0.02 &  10.98 &  0.03 &  6.97 &  0.03 &   18.57 & -0.03 &    9.81 \\ \hline
  \multirow{4}{*}{6} &  (50, 50)  & 0.05 & 0.13 & 0.04 & 0.08 & -0.02 &  27.01 & -0.08 & 17.19 & -0.07 &   25.68 & -0.14 &   13.10 \\
   &  (50, 150)  & 0.05 & 0.09 & 0.02 & 0.06 &  0.06 &  28.59 &  0.06 & 12.45 & -0.07 &    8.70 & -0.11 &    6.01 \\
   &  (150, 50)  & 0.02 & 0.08 & 0.02 & 0.05 &  0.00 &   9.55 & -0.05 &  8.88 &  0.06 &   53.31 &  0.07 &   13.47 \\
   &  (100, 100) & 0.03 & 0.06 & 0.02 & 0.04 &  0.04 &  17.30 & -0.02 &  9.59 &  0.00 &   14.84 & -0.02 &    8.38 \\ \hline
  \multirow{4}{*}{7} &  (50, 50)  & 0.17 & 0.61 & 0.11 & 0.39 &  0.09 &  33.66 &  0.02 & 14.45 & -0.10 &   21.87 & -0.14 &   17.33 \\
   &  (50, 150)  & 0.19 & 0.48 & 0.10 & 0.30 &  0.01 &  25.25 &  0.10 &  6.54 & -0.01 &    8.87 & -0.07 &    7.44 \\
   &  (150, 50)  & 0.06 & 0.28 & 0.06 & 0.19 &  0.00 &   5.25 & -0.05 &  3.62 & -0.05 &   21.97 &  0.00 &   10.64 \\
   &  (100, 100)  & 0.09 & 0.26 & 0.05 & 0.17 & -0.02 &   7.74 &  0.02 &  4.52 &  0.07 &   18.22 & -0.02 &   10.97 \\ \hline
  \multirow{4}{*}{8} &  (50, 50)  & 0.10 & 0.37 & 0.06 & 0.27 &  0.02 &  33.28 &  0.04 &  9.29 & -0.14 &  190.88 & -0.41 &  160.50 \\
   &  (50, 150)  & 0.09 & 0.24 & 0.05 & 0.17 & -0.01 &  27.55 &  0.15 &  6.82 &  0.11 &  106.76 & -0.03 &   97.86 \\
   &  (150, 50)  & 0.03 & 0.22 & 0.02 & 0.15 & -0.05 &   8.84 & -0.05 &  4.72 & -0.12 &  173.09 & -0.35 &  109.84 \\
   &  (100, 100)  & 0.04 & 0.17 & 0.03 & 0.12 &  0.02 &  18.88 &  0.00 &  3.54 & -0.12 &  211.21 & -0.22 &  196.79 \\ \hline
  \multirow{4}{*}{9} &  (50, 50)  & 0.37 & 2.66 & 0.26 & 2.09 &  0.12 &  37.11 &  0.16 &  8.87 &  0.13 &  389.19 & -0.11 &  352.51 \\
   &  (50, 150)  & 0.36 & 1.82 & 0.23 & 1.36 & -0.08 &   7.39 &  0.19 &  3.65 &  0.06 &  135.93 & -0.10 &  128.85 \\
   &  (150, 50)  & 0.13 & 1.16 & 0.10 & 0.90 & -0.01 &   4.89 &  0.03 &  3.01 &  0.23 &  381.40 & -0.08 &  279.96 \\
   &  (100, 100)  & 0.17 & 0.96 & 0.11 & 0.74 & -0.01 &   6.76 &  0.04 &  1.85 & -0.19 &  124.13 & -0.34 &  114.59 \\ \hline
  \multirow{4}{*}{10} &  (50, 50)  & 0.12 & 0.76 & 0.08 & 0.62 & -0.04 &  18.98 &  0.13 &  7.76 &  0.12 & 1461.27 & -0.47 & 1334.57 \\
   &  (50, 150)  & 0.13 & 0.45 & 0.08 & 0.35 & -0.03 &  22.09 &  0.16 &  4.31 & -0.09 &  356.85 & -0.29 &  345.40 \\
   &  (150, 50)  & 0.03 & 0.48 & 0.02 & 0.40 & -0.03 &   9.00 &  0.03 &  3.78 & -0.37 &  743.33 & -0.95 &  640.08 \\
   &  (100, 100)  & 0.06 & 0.33 & 0.04 & 0.28 &  0.00 &  14.03 &  0.09 &  3.61 & -0.04 &  654.22 & -0.34 &  603.20 \\
   \hline
\end{tabular}}
\label{tabratio}
\end{table}

As we can see from Table \ref{tabratio} that the biases  of $\hat\delta
$ and $\tilde\delta$ are quite negligible
in all the cases, and the
proposed estimator $\hat\delta$ has smaller biases in most cases.
Moreover, the proposed estimator $\hat\delta$ outperforms $\tilde\delta$ in terms of MSE.
This is as expected since $\hat\delta
$ uses more information to estimate the population means $\mu_0$ and $\mu_1$.

{For all settings considered in Table \ref{tabratio},  the biases  of $(\hat\sigma_0^2,\hat\sigma_1^2)$ and $(\tilde\sigma_0^2,\tilde\sigma_1^2)$ are quite small.
Meanwhile, the MSEs of  $(\hat\sigma_0^2,\hat\sigma_1^2)$ are significantly smaller than those of $(\tilde\sigma_0^2,\tilde\sigma_1^2)$.
In some settings, such as Model 8 with sample sizes $(n_0,n_1)=(100,100)$, the MSE of $\hat\sigma_0^2$ is smaller than 20\% of the MSE of $\tilde\sigma_0^2$.}

\subsection{Comparison for confidence intervals}
We now examine the finite sample behaviour of the following 95\% CIs of the mean ratio $\delta$:
\begin{enumerate}
\item[--]$\mathcal{I}_{1}$: {Wald-type CI} based on $\log\tilde\delta$ using the quantile of $N(0,1)$;
\item[--]$\mathcal{I}_{1B}$: {bootstrap Wald-type CI} based on $\log\tilde\delta$ using the quantile from the nonparametric bootstrap method;
\item[--]$\mathcal{I}_2$: ELR-based CI proposed by \cite{Wu2012} using the quantile of $\chi^2_1$ distribution;
\item[--]$\mathcal{I}_{2B}$: {bootstrap} ELR-based CI proposed by \cite{Wu2012} using the quantile from the nonparametric bootstrap method;
\item[--]$\mathcal{I}_{3}$: ELR-based CI under the DRM \eqref{drm} proposed by \cite{Wang2018} using the quantile of $\chi^2_1$ distribution;
\item[--] $\mathcal{I}_{4}$: proposed Wald-type CI based on  $\hat\delta$;
\item[--] $\mathcal{I}_{4L}$: proposed Wald-type CI based on $\log\hat\delta$.
 \end{enumerate}

{We note that the normal and $\chi^2_1$ distributions may not provide good approximation to  $\log\tilde\delta$ and the ELR statistic in $\mathcal{I}_2$, respectively, especially when $n$ is not large enough.
This may be because of the specific features in the two-sample semicontinuous data from model \eqref{eq:1}: excessive zeros and severe positive/negative skewness of the positive observations.
Hence, we employ the nonparametric bootstrap method \citep{Efron1993}
to approximate the quantiles of target asymptotic distributions, which leads to $\mathcal{I}_{1B}$ and  $\mathcal{I}_{2B}$.
The number of bootstrap samples is set to be $999$.}

{We construct the first four CIs without using the DRM assumption \eqref{drm} while the rest three CIs are constructed under the DRM.
The performance of a CI is evaluated}
in terms of the coverage probability (CP) and average length
(AL), which are calculated as follows:
\begin{equation*}
    \mbox{CP} (\%) =100\times \frac{\sum_{h=1}^{10000} I(\delta^{(h)}_L<\delta<\delta^{(h)}_U)}{10000},~~~
    \mbox{AL}=\frac{\sum_{h=1}^{10000} \left(\delta^{(h)}_U -\delta^{(h)}_L \right)}{10000}.
\end{equation*}
Here $[\delta^{(h)}_L, \delta^{(h)}_U]$ {denotes a} CI for $\delta$ calculated from the $h$th simulated data.
The simulation results are summarized in Table~\ref{tab:m1}.

\begin{table}[!htt]
\centering
\caption{Coverage probability (\%) and average length of 95\% CIs for $\delta$.}
\tabcolsep 1mm
\scriptsize
{\begin{tabular}{cl | cccc | cccc | cccccc}
\hline
    &  & \multicolumn{2}{c}{$\mathcal{I}_1$} & \multicolumn{2}{c|}{$\mathcal{I}_{1B}$}&
         \multicolumn{2}{c}{$\mathcal{I}_2$}  &
         \multicolumn{2}{c|}{$\mathcal{I}_{2B}$}
         &\multicolumn{2}{c}{$\mathcal{I}_3$} & \multicolumn{2}{c}{$\mathcal{I}_4$}&
         \multicolumn{2}{c}{$\mathcal{I}_{4L}$} \\
  \hline
  Model & $(n_0,n_1)$ & CP & AL & CP & AL & CP & AL & CP & AL & CP & AL & CP & AL & CP & AL\\
  \hline
  \multirow{4}{*}{1} &  (50, 50)  & 92.6 & 1.37 & 94.6 & 1.73 & 91.7 & 1.34 & 94.1 & 1.59 & 94.6 & 1.18 & 93.7 & 1.09 & 94.4 & 1.15 \\
   &  (50, 150)  & 92.6 & 1.09 & 94.0 & 1.26 & 91.8 & 1.08 & 93.7 & 1.22 & 94.9 & 0.94 & 94.2 & 0.90 & 94.8 & 0.93 \\
   &  (150, 50)  & 92.5 & 1.08 & 94.3 & 1.86 & 91.7 & 1.08 & 93.6 & 1.31 & 94.9 & 0.92 & 94.0 & 0.88 & 94.7 & 0.91 \\
   &  (100, 100)  & 93.9 & 0.94 & 95.1 & 1.04 & 93.2 & 0.95 & 94.7 & 1.05 & 94.9 & 0.79 & 94.6 & 0.76 & 95.0 & 0.78 \\ \hline
   \multirow{4}{*}{2}&  (50, 50)  & 91.5 & 2.73 & 94.4 & 3.70 & 89.8 & 2.70 & 93.3 & 3.63 & 94.5 & 2.48 & 91.5 & 2.10 & 94.9 & 2.45 \\
   &  (50, 150)  & 91.7 & 2.13 & 94.0 & 2.65 & 90.7 & 2.16 & 93.6 & 2.82 & 94.5 & 1.99 & 92.9 & 1.74 & 94.9 & 1.94 \\
   &  (150, 50)  & 91.6 & 1.92 & 93.6 & 2.78 & 90.6 & 1.88 & 93.6 & 2.58 & 94.9 & 1.72 & 92.2 & 1.58 & 94.7 & 1.75 \\
   &  (100, 100)  & 92.3 & 1.72 & 94.5 & 2.11 & 92.5 & 1.71 & 94.5 & 2.00 & 94.9 & 1.52 & 93.5 & 1.40 & 95.3 & 1.51 \\ \hline
   \multirow{4}{*}{3}&  (50, 50)  & 92.3 & 1.55 & 94.5 & 1.92 & 91.5 & 1.52 & 94.1 & 1.84 & 94.3 & 1.36 & 92.8 & 1.25 & 94.8 & 1.34 \\
   &  (50, 150)  & 92.6 & 1.17 & 94.3 & 1.36 & 93.0 & 1.15 & 94.7 & 1.30 & 94.8 & 1.04 & 93.5 & 0.97 & 95.0 & 1.01 \\
   &  (150, 50)  & 92.5 & 1.27 & 94.2 & 1.66 & 91.3 & 1.25 & 93.7 & 1.56 & 94.8 & 1.11 & 93.3 & 1.06 & 95.2 & 1.11 \\
   &  (100, 100)  & 94.2 & 1.06 & 95.4 & 1.19 & 92.7 & 1.06 & 93.9 & 1.18 & 94.8 & 0.92 & 94.2 & 0.88 & 95.2 & 0.91 \\ \hline
   \multirow{4}{*}{4}&  (50, 50)  & 91.5 & 2.20 & 94.0 & 2.99 & 90.7 & 2.09 & 93.9 & 2.76 & 93.7 & 1.96 & 90.4 & 1.73 & 94.5 & 1.94 \\
   &  (50, 150)  & 92.6 & 1.60 & 94.4 & 1.88 & 92.0 & 1.60 & 93.8 & 1.86 & 93.9 & 1.47 & 92.2 & 1.35 & 94.3 & 1.45 \\
   &  (150, 50)  & 91.8 & 1.78 & 93.9 & 2.66 & 90.4 & 1.68 & 93.1 & 2.30 & 94.5 & 1.57 & 91.5 & 1.46 & 94.6 & 1.59 \\
   &  (100, 100)  & 93.0 & 1.45 & 94.7 & 1.71 & 92.5 & 1.45 & 94.3 & 1.70 & 94.7 & 1.30 & 92.4 & 1.21 & 94.2 & 1.29 \\ \hline
  \multirow{4}{*}{5} &  (50, 50)  & 92.9 & 2.23 & 95.1 & 2.69 & 91.5 & 2.22 & 93.9 & 2.65 & 95.1 & 1.96 & 93.2 & 1.80 & 94.6 & 1.92 \\
   &  (50, 150)  & 91.8 & 1.88 & 93.4 & 2.21 & 91.0 & 1.86 & 93.4 & 2.18 & 94.7 & 1.69 & 93.8 & 1.57 & 94.7 & 1.65 \\
   &  (150, 50)  & 92.9 & 1.64 & 94.6 & 1.97 & 92.5 & 1.61 & 93.9 & 1.88 & 94.8 & 1.40 & 94.0 & 1.33 & 94.8 & 1.38 \\
   &  (100, 100)  & 93.4 & 1.52 & 94.7 & 1.68 & 93.5 & 1.52 & 94.7 & 1.67 & 94.9 & 1.30 & 94.4 & 1.24 & 95.0 & 1.28 \\ \hline
  \multirow{4}{*}{6} &  (50, 50)  & 91.6 & 3.95 & 94.3 & 5.08 & 90.9 & 3.89 & 93.6 & 5.06 & 94.5 & 3.63 & 92.5 & 3.09 & 94.9 & 3.49 \\
   &  (50, 150)  & 91.4 & 3.31 & 93.6 & 4.18 & 90.5 & 3.32 & 93.3 & 4.45 & 94.8 & 3.13 & 93.2 & 2.75 & 94.9 & 3.03 \\
   &  (150, 50)  & 92.6 & 2.58 & 94.3 & 3.39 & 91.9 & 2.55 & 94.2 & 3.06 & 94.9 & 2.26 & 93.2 & 2.12 & 94.7 & 2.26 \\
   &  (100, 100)  & 92.5 & 2.50 & 94.2 & 2.85 & 92.4 & 2.49 & 94.3 & 2.83 & 94.5 & 2.22 & 94.0 & 2.05 & 94.9 & 2.17 \\ \hline
  \multirow{4}{*}{7} &  (50, 50)  & 92.1 & 2.82 & 94.3 & 3.53 & 91.4 & 2.81 & 94.1 & 3.45 & 94.6 & 2.52 & 92.9 & 2.27 & 95.0 & 2.48 \\
   &  (50, 150)  & 92.3 & 2.37 & 94.1 & 2.86 & 90.9 & 2.38 & 93.2 & 2.90 & 94.6 & 2.18 & 93.8 & 1.99 & 95.0 & 2.13 \\
   &  (150, 50)  & 92.5 & 2.00 & 94.1 & 2.52 & 91.6 & 1.96 & 93.7 & 2.31 & 94.4 & 1.73 & 93.7 & 1.63 & 94.7 & 1.71 \\
   &  (100, 100)  & 93.3 & 1.87 & 94.8 & 2.10 & 92.4 & 1.86 & 93.9 & 2.08 & 94.8 & 1.63 & 94.0 & 1.54 & 95.0 & 1.61 \\ \hline
  \multirow{4}{*}{8} &  (50, 50)  & 92.6 & 2.23 & 94.5 & 2.66 & 91.4 & 2.24 & 93.8 & 2.65 & 94.0 & 1.98 & 92.6 & 1.85 & 94.0 & 1.95 \\
   &  (50, 150)  & 92.6 & 1.79 & 94.2 & 2.07 & 91.5 & 1.77 & 93.5 & 2.00 & 94.8 & 1.61 & 93.8 & 1.53 & 94.5 & 1.58 \\
   &  (150, 50)  & 92.7 & 1.77 & 94.5 & 2.68 & 92.2 & 1.78 & 94.2 & 2.14 & 94.4 & 1.54 & 92.9 & 1.47 & 94.6 & 1.52 \\
   &  (100, 100)  & 93.5 & 1.56 & 95.0 & 1.75 & 92.8 & 1.56 & 94.4 & 1.72 & 94.3 & 1.37 & 93.5 & 1.30 & 94.5 & 1.34 \\ \hline
  \multirow{4}{*}{9} &  (50, 50)  & 91.3 & 5.78 & 93.8 & 7.90 & 90.8 & 5.70 & 94.5 & 7.74 & 93.3 & 5.41 & 89.3 & 4.60 & 93.5 & 5.42 \\
   &  (50, 150)  & 92.0 & 4.56 & 94.0 & 5.72 & 90.9 & 4.63 & 93.6 & 6.49 & 94.2 & 4.45 & 91.8 & 3.86 & 94.0 & 4.33 \\
   &  (150, 50)  & 91.7 & 4.10 & 94.0 & 8.56 & 91.1 & 3.89 & 93.6 & 5.23 & 93.3 & 3.65 & 90.2 & 3.35 & 94.0 & 3.71 \\
   &  (100, 100)  & 92.5 & 3.66 & 94.3 & 4.31 & 91.9 & 3.64 & 94.0 & 4.26 & 93.7 & 3.39 & 91.7 & 3.08 & 94.3 & 3.34 \\ \hline
  \multirow{4}{*}{10} &  (50, 50)  & 92.3 & 3.26 & 94.5 & 4.10 & 91.3 & 3.19 & 94.1 & 4.00 & 93.0 & 3.00 & 89.6 & 2.74 & 93.3 & 3.01 \\
   &  (50, 150)  & 92.8 & 2.46 & 94.5 & 2.83 & 91.8 & 2.42 & 93.8 & 2.75 & 94.1 & 2.33 & 92.6 & 2.16 & 94.0 & 2.28 \\
   &  (150, 50)  & 92.4 & 2.68 & 94.1 & 3.83 & 90.8 & 2.63 & 93.3 & 3.43 & 93.0 & 2.41 & 90.4 & 2.27 & 93.7 & 2.44 \\
   &  (100, 100)  & 93.7 & 2.22 & 95.1 & 2.53 & 92.6 & 2.22 & 94.5 & 2.52 & 94.0 & 2.06 & 91.8 & 1.94 & 93.0 & 2.04 \\
   \hline
\end{tabular}}
\label{tab:m1}
\end{table}

From Table~\ref{tab:m1}, {we observe that} the bootstrap Wald-type CI $\mathcal{I}_{1B}$ and bootstrap ELR-based CI  $\mathcal{I}_{2B}$ have much better coverage accuracy than
$\mathcal{I}_{1}$ and $\mathcal{I}_{2}$, respectively.
Comparing $\mathcal{I}_{1B}$ and $\mathcal{I}_{2B}$, we see that $\mathcal{I}_{1B}$
has slightly more accurate CP in most cases, but $\mathcal{I}_{2B}$ has shorter AL in most cases.
The behaviour of $\mathcal{I}_{3}$ and $\mathcal{I}_{4L}$ are very comparable and satisfactory in terms of both CP and AL in all cases, while $\mathcal{I}_{4}$ gives shorter ALs and experiences lower coverage rates compared with $\mathcal{I}_{3}$ and $\mathcal{I}_{4L}$, especially in the cases with small sample sizes.

In general, the performance of the CIs constructed under the DRM is better than that of the CIs {constructed without using} the DRM assumption.
{In conclusion, the simulation results of $\mathcal{I}_{3}$ and $\mathcal{I}_{4L}$ seem most attractive regarding both CP and AL.
However, the computational time and complexity of the proposed $\mathcal{I}_{4L}$ is far less than that of $\mathcal{I}_{3}$, and thus may be preferred.}


\section{Real data analysis}
\label{realdata}
In this section, we illustrate the performance of our proposed method by analyzing two {real datasets.
Our interest lies in estimating the mean ratio} $\delta$ and  population variances $\sigma^2_0,\sigma^2_1$, and constructing  the CIs for $\delta$.


The first dataset is from a biological study of the seasonal activity patterns of a species of field mice, which is originally taken from \cite{mice}.  It consists of the average distances (in meters) traveled between captures by those mice at least twice in a given month.
A summary of this dataset is presented in Table \ref{micedata}.
\begin{table}[!htt]
   \centering
    \caption{Summary of mice dataset. }
    \tabcolsep 1mm
\scriptsize
    \begin{tabular}{ccc}
    \hline
      Season  &  sample size & proportion (number) of zeros\\
    \hline
       Spring  & 17& 0.176 (3)\\
       Summer & 27 & 0.111 (3)\\
       Autumn & 27 & 0.370 (10)\\
       Winter & 34 &  0.294 (10)\\
      \hline
    \end{tabular}
    \label{micedata}
\end{table}

From Table \ref{micedata}, there are considerable proportions of zero measurements, especially in Fall and Winter.
{\cite{Wang2018} have conducted hypothesis tests to discover if the mean traveled distance differs among the four seasons. They found no significant difference between the mean distance in  Spring and that in Summer.  Hence, we combine the distance measurements in Spring and Summer into one sample and denote it as sample $0$.
Similarly, sample $1$ is obtained by merging the distance measurements in Autumn and Winter.}

To analyze the dataset by our proposed method, we need to choose an appropriate $\bq(x)$ in the DRM \eqref{drm}.
To balance the model fitting and model complexity, we choose $\bq(x) = \log(x)$.
We apply the goodness-of-fit test proposed by \cite{Qin1997} for this choice to the mice data,
which  gives $p$-value $=0.64$. This may indicate that $\bq(x) = \log(x)$ is suitable for this dataset.

{All the methods discussed in our simulation studies are applied here. Our proposed estimate $\hat\delta = 0.487$ while the fully nonparametric estimate $\tilde \delta = 0.483$.
The proposed semiparametric estimates of the two-sample variances are $\hat\sigma^2_0 = 869.583$ for sample $0$ and $\hat\sigma^2_1 = 268.774$ for sample $1$. As a comparison, their fully nonparametric estimates are $\tilde\sigma^2_0 = 932.966$ for sample $0$ and $\tilde\sigma^2_1 = 239.961$ for sample $1$.}
Based on the simulation results in Table \ref{tabratio}, the proposed point estimates are expected to be more accurate.

The results of 95\% CIs for $\delta$ are presented in Table \ref{miceCI}.
{Among all the considered CIs, $\mathcal{I}_1$ is the shortest and  $\mathcal{I}_{1B}$ has the longest interval length.
The lower and upper bounds of $\mathcal{I}_4$ tend to be smaller than those of other CIs.
The results of $\mathcal{I}_{2B}$, $\mathcal{I}_3$, and $\mathcal{I}_{4L}$ are similar.
All these CIs do not include 1, which indicates a significant mean difference between the two samples.}

\begin{table}[!htt]
\centering
\caption{95\% CIs of $\delta$ in mice data. }
    \tabcolsep 1mm
\scriptsize
\begin{tabular}{cccccccc}
\hline
& $\mathcal{I}_1$ & $\mathcal{I}_{1B}$ & $\mathcal{I}_{2}$ & $\mathcal{I}_{2B}$ & $\mathcal{I}_{3}$ & $\mathcal{I}_{4}$&$\mathcal{I}_{4L}$\\
\hline
Lower bound & 0.341 & 0.314 & 0.318 & 0.322 & 0.325 &0.295 & 0.328 \\
Upper bound & 0.682 & 0.741 & 0.726 & 0.716 & 0.721 &0.679 & 0.722 \\
Length & 0.341 & 0.427 & 0.408 & 0.393 & 0.396 & 0.383 & 0.393 \\
\hline
\end{tabular}
\label{miceCI}
\end{table}

The second dataset is about the  methylation of DNA,
which is a common method of gene regulation. The methylation patterns of tumor cells can be compared to those of normal cells; moreover, there are also differences between different types of cancer.
The DNA methylation can served as a biomarker of diagnosing cancers.
The dataset, collected from \cite{neuhauser2011nonparametric}, consists of two samples of methylation measurements: small cell lung cancer (sample $0$) and non-small cell lung cancer (sample $1$).
When methylation is undetectable or is partially present, the result of measurement is negative, which is treated as a zero value. The fully presence of methylation gives a positive value.
Sample $0$ contains 41 measurements, out of which 25 are zero values. There are 46 measurements in sample $1$ and 16 of them are zero values.

\cite{satter2020jackknife} argued that this dataset is highly skewed. This may suggest that the dataset can be well fitted by the DRM with $\bq(x) = \log(x)$.
The goodness-of-fit test of \cite{Qin1997}  gives  $p$-value $=0.133$. Therefore, there is no strong evidence to reject the DRM with $\bq(x) = \log(x)$.

{All the methods discussed in our simulation studies are applied here.
Our proposed estimate $\hat\delta = 2.906$ while the fully nonparametric estimate $\tilde \delta = 3.679$.
For the two-sample variances, our proposed semiparametric estimates are $\hat\sigma^2_0 = 388.562$  for sample $0$ and $\hat\sigma^2_1 = 1028.079$ for sample $1$, while the fully nonparametric estimates are $\tilde\sigma^2_0 = 406.796$ for sample $0$ and $\tilde\sigma^2_1 = 1017.072$  for sample $1$.
There are some differences between the proposed estimates and the fully nonparametric estimates, especially for estimating $\delta$.
We rely more on our proposed estimates since the performance of the proposed estimators is more accurate than other competitors as observed in our simulations.
}

Table \ref{DNACI} presents the results of 95\% CIs for $\delta$.
According to the simulation results in Table \ref{tab:m1}, the CIs $\mathcal{I}_{1B}$, $\mathcal{I}_{2B}$,  $\mathcal{I}_3$, and $\mathcal{I}_{4L}$
are more trustable in terms of coverage accuracy.
Among these four CIs, the CIs $\mathcal{I}_{1B}$ and $\mathcal{I}_{2B}$ contain 1,
whereas $\mathcal{I}_3$ and $\mathcal{I}_{4L}$ do not contain 1.
This indicates that $\mathcal{I}_3$ and $\mathcal{I}_{4L}$ provide more evidence than $\mathcal{I}_{1B}$ and $\mathcal{I}_{2B}$
to reject $H_0:\delta=1$.
{Comparing between $\mathcal{I}_3$ and $\mathcal{I}_{4L}$, $\mathcal{I}_{4L}$ has slightly shorter interval length.}

\begin{table}[!htt]
\centering
\caption{95\% CIs of $\delta$ in methylation data. }
    \tabcolsep 1mm
\scriptsize
\begin{tabular}{cccccccc}
\hline
& $\mathcal{I}_1$ & $\mathcal{I}_{1B}$ & $\mathcal{I}_{2}$ & $\mathcal{I}_{2B}$ & $\mathcal{I}_{3}$ & $\mathcal{I}_{4}$&$\mathcal{I}_{4L}$\\
\hline
Lower bound &  1.291 &  0.650 &  1.158 &  0.568 &  1.278 &  0.362 &  1.211 \\
  Upper bound & 10.485 & 20.838 & 12.306 & 27.631 &  7.527 &  5.451 &  6.975 \\
  Length &  9.194 & 20.189 & 11.148 & 27.063 &  6.249 &  5.089 &  5.764 \\
   \hline
\end{tabular}
\label{DNACI}
\end{table}

\section{Concluding remarks}
\label{conclude}

{In this paper,  we propose new statistical procedures for making semiparametric inference on  the general parameters $\bpsi$ defined in (\ref{psi}) and their functions
${\bf g}(\bpsi)$ with two samples of  semicontinuous observations.}
The parameters $\bpsi$ include the linear functionals $\bpsi_0$ and $\bpsi_1$ as special cases.
Under the semiparametric DRM \eqref{drm},
we construct the MELEs of $\bpsi$ and establish the asympotic normality of the MELEs of  $\bpsi$.
The MELEs of $\bpsi_0$ and $\bpsi_1$ are shown to be more efficient than
{the fully nonparametric alternatives in both theory and simulations.
We further apply the developed asymptotic results to construct the confidence regions and perform hypothesis tests for $\bpsi$ and ${\bf g}(\bpsi)$.
It is worth mentioning that the proposed methods and the general results can be applied to many important summary quantities, such as the uncentered and centered moments, the mean ratio, the coefficient of variation, and the generalized entropy class of inequality measures.
As illustration, we consider the construction of CI for the mean ratio of two such populations.
Simulation results show that the proposed Wald-type CIs have similar performance as the ELR-based CI under the DRM with less computational cost.}
We have implemented our methods in \texttt{R} language. The \texttt{R} code is available upon request.

{With the observable advantages in making semiparametric inference on the linear functions,}
it would be interesting to extend the current framework to
general expectation functionals and their functions, for example, the receiver operating characteristic (ROC) curve, the area under the ROC curve, and
Gini index.
The theoretical development may become even complicated and challenging.
We leave them as future research.

\section*{Acknowlegement}
The authors would like to thank Dr. Changbao Wu for his constructive and helpful comments.
Dr. Wang's work is supported in part by National Natural Science Foundation of China grant numbers 12001454, 11971404, 71988101, and Natural Science Foundation of Fujian Province grant number 2020J01031.
Dr. Li's work is supported in part by the Natural Sciences and Engineering Research Council of Canada grant number RGPIN-2020-04964.

\medskip 

\newpage

{\centering {\large {\bf Supplementary material for \\
``Semiparametric inference on general functionals of two semicontinuous  populations''}}\par }
\bigskip

\no
This document supplements the paper entitled ``Semiparametric inference on general functionals of two semicontinuous populations''.
It contains proofs for Theorem 1,  Theorem  2, and Corollary 1.
Section 1 introduces some notations and makes some preparations.
Section 2 presents some useful lemmas.
Our proofs of Theorem 1, Theorem 2, and Corollary 1 are given in Sections 3, 4, and 5, respectively.

\label{appendix}
\setcounter{section}{0}
\section{Some preparations}
Recall that \begin{equation*}\label{neweq:1}
  X_{i1},\cdots, X_{in_i} \sim F_i(x)= v_iI(x \geq 0)+(1-v_i)I(x>0)G_i(x), ~~\mbox{for~}~i=0,1,
\end{equation*}
where $v_i\in (0,1)$, $n_i$ is the sample size for sample $i$, $I(\cdot)$ is an indicator function, and $G_i(\cdot)$'s are cumulative distribution functions (CDFs) of positive observations in sample $i$.
We link $G_0(x)$ and $G_1(x)$ via a density ratio model (DRM):
\begin{equation}
\label{newdrm}
dG_1(x) = \exp\{\alpha +\boldsymbol{\beta}^\top\bq(x)\}dG_0(x),
\end{equation}
where $\bq(x)$ is a pre-specified, non-trivial, $d$-dimensional basis function.

Recall that we denote $n_{i0}$ and $n_{i1}$ as the (random) number of zero observations and positive observations, respectively, in each sample $i=0,1$.  Clearly $n_i = n_{i0}+n_{i1}$, for $i=0,1$.
Without loss of generality, we assume that the first $n_{i1}$ observations in group $i$, $X_{i1},\cdots, X_{in_{i1}}$, are positive,
and the rest $n_{i0}$ observations are 0.
We use $n$ to denote the total (fixed) sample size, i.e., $n=n_0+n_1$.

We let $\bnu = (\nu_0,\nu_1)^\top$ and $\btheta = (\alpha, \bbeta^\top)^\top$.
The maximum empirical likelihood estimators (MELEs) of $\bnu$ and $\btheta$ respectively maximize
$
\ell_0\left(\bnu\right)
$
and
$
\ell_1(\btheta)
$, where
$$
\ell_0\left(\bnu\right) = \sum_{i=0}^1\log\left\{ v_i^{n_{i0}}\left( 1-v_i\right)^{n_{i1}} \right\}
$$
and
$$
\ell_1(\btheta)=
- \Sum
  \log\left\{ 1 + \hat\rho[\exp\{\alpha+ \bbeta^{\top} \bq(X_{ij})\}-1] \right\}
+ \sum_{j=1}^{n_{11}} \{\alpha+ \bbeta^{\top}  \bq(X_{1j}) \}
$$
with $\hat\rho = {n_{11}}/{(n_{01}+n_{11})}$ being a random variable.
That is,
\begin{equation}
\label{mele1}
\hat\nu=\arg\max_{\bsnu} \ell_0\left(\bnu\right)
~~
\mbox{ and }
~~
\hat\btheta=\arg\max_{\bstheta}\ell_1(\btheta).
\end{equation}
Note that
\begin{equation}
\label{mele2}
\sum_{i=0}^1\sum_{j=1}^{n_{i1}}
\frac{1}{n_{01}+n_{11}}\frac{1} {1 + \hat{\rho}[\exp\{\hat\alpha + \hat\bbeta^{\top} \bq(X_{ij})\}-1]  }=1,
\end{equation}
which ensures that the MELE of $G_0(x)$ is a CDF.

For the convenience of presentation, we recall and introduce some more notations.
We use $\bnu^{*}$ and $\btheta^{*}$ to denote the true values of $\bnu$ and $\btheta$, respectively.
Let  $\bQ(x) = (1,\bq(x)^\top)^\top$ and
\bas
&w = n_0/n,~\Delta^*=w(1-\nu_0^*)+(1-w)(1-\nu_1^*),
~\rho^*=\frac{(1-w)(1-\nu_1^*)}{\Delta^*},\\
&\omega(x;\btheta)=\exp\{\btheta^{\top} \bQ(x)\},~\omega(x) = \omega(x;\btheta^*),\\
&h(x)=1+\rho^* \{\omega(x)-1\},~h_1(x)=\rho^* \omega(x)/h(x),~h_0(x) =(1-\rho^*)/h(x).
\eas
Note that $\omega(\cdot)$, $h(\cdot)$, $h_0(\cdot)$, and $h_1(\cdot)$ depend on $\btheta^*$ and/or $\rho^*$ and $h_0(x) + h_1(x) =1$. We use $\sum_{ij}$ to denote summation over the full range of data in the  rest of supplementary material.

Further define $\hat\boeta=(\hat\bnu^\top,\hat\rho,\hat\btheta^\top)^\top$  and $\boeta^*=(\bnu^{*\top},\rho^*,\btheta^{*\top})^\top$.
To derive the asymptotic properties, we define an expanded function:
\ba
\nonumber
H(\bnu,\rho,\btheta)
&=&
n_{00}\log(\nu_0)+n_{01}\log(1-\nu_0)+ n_{10}\log(\nu_1)+n_{11}\log(1-\nu_1)\\
&&
- \Sum \log\left\{ 1 + \rho[\exp\{\btheta^{\top} \bQ(X_{ij})\}-1] \right\}
+ \sum_{j=1}^{n_{11}} \{\bbeta^{\top}  \bq(X_{1j}) \}.
\label{H.fun}
\ea
By (\ref{mele1}), we get
\begin{equation}
\label{mele11}
\frac{\partial H(\hat\bnu,\hat\rho,\hat\btheta)}{\partial \bnu}=\0\mbox{ and }
\frac{\partial H(\hat\bnu,\hat\rho,\hat\btheta)}{\partial \btheta}=\0.
\end{equation}
From (\ref{mele2}), we can verify that
\begin{equation}
\label{mele21}
\frac{\partial H(\hat\bnu,\hat\rho,\hat\btheta)}{\partial \rho}=0.
\end{equation}
Then (\ref{mele11}) and (\ref{mele21})
together imply  that $\hat\boeta $ satisfies
\begin{equation}
\frac{\partial H(\hat\boeta)}{\partial \boeta}=\0,
\end{equation}
which serves as the starting point of our proof for $\hat \boeta$.

Next, we apply the first-order Taylor expansion to $\partial H(\hat\boeta)/\partial \boeta$ to find an approximation for
$\hat \boeta$. In this process,  the first and second derivatives of $H(\bnu,\rho,\btheta)$ play important roles.
Their detailed forms are given below.

%
%
\subsection{First derivatives of $H(\bnu,\rho,\btheta)$}

After some calculations, we  find the first derivatives of $H(\bnu,\btheta,\rho)$ as follows:
\bas
\frac{\partial H(\bnu,\rho,\btheta)}{\partial \bnu}
&=&
\left(\frac{\partial H(\bnu,\rho,\btheta)}{\partial \nu_0},\frac{\partial H(\bnu,\rho,\btheta)}{\partial \nu_1}\right)^\top
=
\left(\frac{n_{00}}{\nu_0}-\frac{n_{01}}{1-\nu_0}, \frac{n_{10}}{\nu_{1}}-\frac{n_{11}}{1-\nu_1}\right)^\top,
\\
\frac{\partial H(\bnu,\rho,\btheta)}{\partial \rho}
&=&
- \sum_{ij}\frac{\omega(X_{ij};\btheta)-1}{1+\rho\{\omega(X_{ij};\btheta)-1\}} I(X_{ij}>0),
\\
\frac{\partial H(\bnu,\rho,\btheta)}{\partial \btheta}
&=&
\sum_{j=1}^{n_1}\bQ(X_{1j})I(X_{1j}>0) - \sum_{ij}\frac{\rho \omega(X_{ij};\btheta)}{1+\rho\{\omega(X_{ij};\btheta)-1\} } \bQ(X_{ij}) I(X_{ij}>0).
\eas

We evaluate the above derivatives at  $\boeta^*$ and define
\ba
\label{bsn}
\bS_n=\frac{\partial H(\boeta^*)}{\partial \boeta}
=\left( \begin{array}{c}
 \frac{\partial H(\boeta^*)}{\partial \bsnu} \\
 \frac{\partial H(\boeta^*)}{\partial \rho} \\
 \frac{\partial H(\boeta^*)}{\partial \bstheta}
  \end{array}\right)
=\left( \begin{array}{c}
 \bS_{n,\bsnu} \\
 S_{n,\rho} \\
 \bS_{n,\bstheta}
  \end{array}\right),
\ea
where the corresponding entries are
\bas
\bS_{n,\bsnu}
&=& \left( \frac{n_{00}}{\nu_{0}^*}-\frac{n_{01}}{1-\nu_0^*}, \frac{n_{10}}{\nu_{1}^*}-\frac{n_{11}}{1-\nu_1^*} \right)^\top,
\\
S_{n,\rho}
&=& -\sum_{ij}\frac{\omega(X_{ij})-1}{h(X_{ij})} I(X_{ij}>0),
\\
\bS_{n,\bstheta}
&=& \sum_{j=1}^{n_1}\bQ(X_{1j})I(X_{1j}>0) - \sum_{ij} h_1(X_{ij})\bQ(X_{ij}) I(X_{ij}>0).
\eas

\subsection{Second derivatives of $H(\bnu,\rho,\btheta)$}

We next calculate the second derivatives of $H(\bnu,\rho,\btheta)$ and evaluate them at $\boeta^*$.
This leads to
\ba
\label{H.2nd}
\frac{\partial^2 H(\boeta^*)}{\partial \boeta \partial \boeta^\top}
=\left( \begin{array}{ccc}
 \frac{\partial^2 H(\boeta^*)}{\partial \bsnu \partial \bsnu^\top} &
 \frac{\partial^2 H(\boeta^*)}{\partial \bsnu \partial \rho} &
 \frac{\partial^2 H(\boeta^*)}{\partial \bsnu \partial \bstheta^\top} \\
 \frac{\partial^2 H(\boeta^*)}{\partial \rho \partial \bsnu^\top} &
 \frac{\partial^2 H(\boeta^*)}{\partial \rho^2} &
 \frac{\partial^2 H(\boeta^*)}{\partial \rho \partial \bstheta^\top} \\
 \frac{\partial^2 H(\boeta^*)}{\partial \bstheta \partial \bsnu^\top} &
 \frac{\partial^2 H(\boeta^*)}{\partial \bstheta \partial \rho} &
 \frac{\partial^2 H(\boeta^*)}{\partial \bstheta \partial \bstheta^\top} \\
  \end{array}\right),
\ea
where
\bas
\frac{\partial^2 H(\boeta^*)}{\partial \bnu \partial \bnu^\top}
&=&
\diag\left\{ -\frac{n_{00}}{\nu_{0}^{*2}}-\frac{n_{01}}{(1-\nu_0^*)^2}, -\frac{n_{10}}{\nu_{1}^{*2}}-\frac{n_{11}}{(1-\nu_1^*)^2} \right\} ,
\\
\frac{\partial^2 H(\boeta^*)}{\partial \rho^2}
&=&
 -\sum_{ij}\frac{ - \{\omega(X_{ij})-1\}^2}{h(X_{ij})^2} I(X_{ij}>0) ,
\\
\frac{\partial^2 H(\boeta^*)}{\partial \bnu \partial \rho}
&=&
\left\{\frac{\partial^2 H(\boeta^*)}{\partial \rho \partial \bnu^\top}\right\}^\top
= \0 ,
\\
\frac{\partial^2 H(\boeta^*)}{\partial \btheta \partial \rho}
&=&
\left\{\frac{\partial^2 H(\boeta^*)}{\partial \rho \partial \btheta^\top}\right\}^\top
=
-\sum_{ij}\frac{ \omega(X_{ij})}{h(X_{ij})^2} \bQ(X_{ij}) I(X_{ij}>0), 
\\
\frac{\partial^2 H(\boeta^*)}{\partial \btheta \partial \btheta^\top}
&=&
 -\sum_{ij} h_0(X_{ij}) h_1(X_{ij}) \{\bQ(X_{ij})\bQ(X_{ij})^\top\} I(X_{ij}>0),
\\
\frac{\partial^2 H(\boeta^*)}{\partial \bnu \partial \btheta^\top}
&=&
\left\{\frac{\partial^2 H(\boeta^*)}{\partial \btheta \partial \bnu^\top}\right\}^\top = \0.
\eas

\section{Some useful lemmas}

In the proof of Theorem 1, we need the expectation of
${\partial^2 H(\boeta^*)}/(\partial \boeta \partial \boeta^\top)$
and the asymptotic property of $\bS_n$.
The following lemma is used to ease the calculation burden in our later proofs.

\begin{lemma}
\label{lem2}
Suppose that $f$ is an arbitrary vector-valued function.
Let $E_0(\cdot)$ represents the expectation with respect to $G_0$ and $X$ refers to a random variable from $G_0$.
Then \\
\[
E \left\{\sum_{ij}f(X_{ij}) I(X_{ij}>0) \right\} = n\Delta^* E_0\{h(X)f(X)\} .
\]
\end{lemma}

\proof
Note that
\bas
E\left\{\sum_{ij} f(X_{ij}) I(X_{ij}>0) \right\}
&=&\sum_{i=0}^1 n_{i}{ E}\{f(X_{i1}) I(X_{i1}>0) \}  \\
&=&n_{0}(1-\nu_0^*) E_0\{f(X) \} + n_{1}(1-\nu_1^*) E_0\{\omega(X)f(X) \},
\eas
where we use the DRM (\ref{newdrm}) in the last step.
Using the facts that $w=n_0/n$ and $1-w=n_1/n$, we further have
\bas
E\left\{\sum_{ij} f(X_{ij}) I(X_{ij}>0) \right\}
&=&n w(1-\nu_0^*) E_0\{f(X) \} + n(1-w)(1-\nu_1^*) E_0\{\omega(X)f(X) \}.
\eas
Recall the definitions of $\Delta^*$ and $\rho^*$.
Then we get
\bas
E\left\{\sum_{ij} f(X_{ij}) I(X_{ij}>0) \right\}
&=&n\Delta^*E_0\{(1-\rho^*)f(X) \}+n\Delta^*E_0[\rho^* \omega(X)f(X)]  \\
&=&n\Delta^* E_0\{h(X)f(X) \} .
\eas
This finishes the proof.
$\hfill \square$

\smallskip

With the help of Lemma \ref{lem2}, we  calculate the expectation of
$\partial^2 H(\boeta^*)/(\partial \boeta \partial \boeta^\top)$.

\begin{lemma}
\label{lem.A}
With the form of $\partial^2 H(\boeta^*)/(\partial \boeta \partial \boeta^\top)$ given in \eqref{H.2nd},
we have
\bas
-\frac{1}{n}{ E}\left\{\frac{\partial^2 H(\boeta^*)}{\partial \boeta \partial \boeta^\top}\right\}
=\bA=\left( \begin{array}{ccc}
 \bA_{\bsnu} & \0 & \0 \\
 \0 & -A_{\rho} & \bA_{\rho,\bstheta} \\
 \0 & \bA_{\bstheta,\rho} & \bA_{\bstheta} \\
  \end{array}\right),
\eas
where
\bas
\bA_{\bsnu} &=&
\diag\left\{ \frac{w}{\nu_0^*(1-\nu_0^*)}, \frac{1-w}{\nu_1^*(1-\nu_1^*)} \right\},~
\bA_{\bstheta} =
\Delta^*(1-\rho^*)E_0\left[h_1(X) \bQ(X)\bQ^\top(X) \right],\\
A_{\rho}
&=& \Delta^*E_0\left\{ \frac{ \{\omega(X)-1\}^2}{h(X)} \right\} = \{\rho^*(1-\rho^*)\}^{-1}\left[\Delta^*-\{\rho^*(1-\rho^*)\}^{-1}\bfe^\top \bA_{\bstheta}\bfe\right], \\
\bA_{\bstheta,\rho}
&=& \bA_{\rho,\bstheta}^\top
= \Delta^*E_0\left\{ \frac{\omega(X)}{h(X)} \bQ(X) \right\} = \{\rho^*(1-\rho^*)\}^{-1} \bA_{\bstheta}\bfe
\eas
with $\bfe=(1,\0_{d\times1}^\top)^\top$.
\end{lemma}
\proof
Note that $n_{00}\sim {\rm Bin}(n_0,\nu_0)$ and $n_{10}\sim {\rm Bin}(n_1,\nu_1)$, where ``${\rm Bin}$" denotes the binomial distribution.
With the facts $w=n_0/n$ and $1-w=n_1/n$,
 it can be easily show that
$$
-\frac{1}{n}{ E}\left\{\frac{\partial^2 H(\boeta^*)}{\partial \nu \partial \nu^\top}\right\}
={\bA}_{\bsnu}.
$$

Next, we apply Lemma \ref{lem2} to find the reminging entries of $E\left\{ \partial^2 H(\boeta^*)/(\partial \boeta \partial \boeta^\top)\right\}$.
We use
$$
E\left\{ \frac{\partial^2 H(\boeta^*)} {\partial \btheta \partial \btheta^\top} \right\}$$ as illustration.
For other entries, the idea is similar and hence we omit the details.

Note that
\bas
-\frac{1}{n}E\left\{ \frac{\partial^2 H(\boeta^*)}{\partial \btheta \partial \btheta^\top}\right\}
&=& \frac{1}{n}E\left\{\sum_{ij} h_0(X_{ij}) h_1(X_{ij})\bQ(X_{ij})\bQ(X_{ij})^\top I(X_{ij}>0) \right\}\\
&=& \Delta^*E_0\left\{h(X) h_0(X) h_1(X) \bQ(X)\bQ(X)^\top\right\}\\
&=& \Delta^*(1-\rho^*)E_0\left\{h_1(X) \bQ(X)\bQ(X)^\top\right\},
\eas
where we have used Lemma \ref{lem2} in the second step, and used the fact that $h(x) h_0(x) =1-\rho^*$ in the third step.
This finishes the proof.
$\hfill \square$

We now study the asymptotic properties of $\bS_n$ defined in \eqref{bsn}. Recall
$\bW = \left( (1-\nu_0^*)^{-1}, -(1-\nu_1^*)^{-1} \right)$ and define $S= w^{-1} + (1-w)^{-1}$.
\begin{lemma}
\label{lem.Sn}
With the form of $\bS_n$ in\eqref{bsn}, we have, as $n\to\infty$
\[
n^{-1/2}\bS_n \to N(\0, \bB)
\]
in distribution,
where
\bas
\bB
&=& \left( \begin{array}{ccc}
 \bA_{\bsnu} & \0 & \0 \\
 \0 & A_{\rho} & \0 \\
 \0 & \0 & \bA_{\bstheta} \\
  \end{array}\right)
+
\left( \begin{array}{ccc}
 \0 & -\rho^*(1-\rho^*) A_{\rho} \bW^\top & \bW^\top\bfe^\top \bA_{\bstheta} \\
 -\rho^*(1-\rho^*) A_{\rho}\bW & -S \{\rho^{*}(1-\rho^*)\}^2 A_{\rho}^2 & S \rho^*(1-\rho^*) A_{\rho}\bfe^\top \bA_{\bstheta} \\
  \bA_{\bstheta}\bfe \bW & S \rho^*(1-\rho^*) A_{\rho}\bA_{\bstheta}\bfe & -S \bA_{\bstheta}\bfe (\bA_{\bstheta}\bfe)^\top \\
  \end{array}\right).
\eas
\end{lemma}

\proof
Using the results in Lemma \ref{lem2},
it is easy to show that $E(\bS_n)=\0$.
We omit the details.

Next, we verify that  $\var(\bS_n)=\bB$.
For convenience, we write $\bB$ as
\[
\bB
=\left( \begin{array}{ccc}
 \bB_{11} & \bB_{12} & \bB_{13} \\
 \bB_{21} & B_{22} & \bB_{23} \\
 \bB_{31} & \bB_{32} & \bB_{33} \\
  \end{array}\right).
\]
In the following,  we concentrate on deriving $\bB_{13}$. Other entries can be similarly obtained and hence we omit the details.

Note that $\bS_{n,\bsnu}$ and $\bS_{n,\bstheta}$  can be rewritten as
\bas
\bS_{n,\nu_0} &=& \frac{n_{00}}{\nu_0^*} - \frac{n_{01}}{1 -\nu_0^*}= -\frac{n_{01}}{\nu_0^*(1-\nu_0^*)}=-\frac{1}{\nu_0^*(1-\nu_0^*)}\sum_{j=1}^{n_{0}} I(X_{0j}>0),\\
\bS_{n,\nu_1} &=& \frac{n_{10}}{\nu_1^*} - \frac{n_{11}}{1 -\nu_1^*}= -\frac{n_{11}}{\nu_1^*(1-\nu_1^*)}=
-\frac{1}{\nu_1^*(1-\nu_1^*)} \sum_{j=1}^{n_1}I(X_{1j}>0),\\
\bS_{n,\bstheta} &=& \sum_{j=1}^{n_1}\bQ(X_{1j})I(X_{1j}>0) - \sum_{ij} h_1(X_{ij})\bQ(X_{ij}) I(X_{ij}>0)\\
&=& \sum_{j=1}^{n_1}h_0(X_{1j})\bQ(X_{1j})I(X_{1j}>0) - \sum_{j=1}^{n_0}h_1(X_{0j})\bQ(X_{0j})I(X_{0j}>0).
\eas
Then we have
\bas
\frac{1}{n}\cov(\bS_{n,\nu_0},\bS_{n,\bstheta}^\top) &=& \frac{1}{n \nu_0^*(1-\nu_0^*)}\cov\left\{\sum_{j=1}^{n_{0}} I(X_{0j}>0), \sum_{j=1}^{n_0}h_1(X_{0j})\bQ(X_{0j})^\top I(X_{0j}>0) \right\}
\\
&=& \frac{n_0}{n \nu_0^*(1-\nu_0^*)} \left[ (1-\nu_0^*)E_0\left\{ h_1(X) \bQ(X)^\top \right\} - (1-\nu_0^*)^2E_0\left\{ h_1(X) \bQ(X)^\top \right\} \right] \\
&=& w E_0\left\{ h_1(X) \bQ(X)^\top \right\} \\
&=& (1-\nu_0^*)^{-1} (\bA_{\bstheta}\bfe)^\top.
\eas
Similarly,
\bas
&&\frac{1}{n}\cov(\bS_{n,\nu_1},\bS_{n,\bstheta}^\top)\\
&=& \frac{-1}{n \nu_1^*(1-\nu_1^*)}\cov\left\{\sum_{j=1}^{n_{1}} I(X_{1j}>0), \sum_{j=1}^{n_1}h_0(X_{1j})\bQ(X_{1j})^\top I(X_{1j}>0) \right\}
\\
&=& \frac{-n_1}{n \nu_1^*(1-\nu_1^*)} \left[ (1-\nu_1^*)E_0\left\{ h_0(X)\omega(X) \bQ(X)^\top \right\} - (1-\nu_1^*)^2E_0\left\{ h_0(X)\omega(X) \bQ(X)^\top \right\} \right] \\
&=& -(1-w) E_0\left\{ \omega(X)h_0(X) \bQ(X)^\top \right\} \\
&=& -(1-w)\cdot \frac{1-\rho^*}{\rho^*} E_0\left\{ h_1(X) \bQ(X)^\top \right\} \\
&=& -(1-\nu_1^*)^{-1} (\bA_{\bstheta}\bfe)^\top.
\eas
Recall $\bW = \left( (1-\nu_0^*)^{-1}, -(1-\nu_1^*)^{-1} \right)$.
Then $\bB_{13} = \bW^\top\bfe^\top\bA_{\bstheta}$.

Note that  $\bS_n$ in \eqref{bsn} is a sum of independent random vectors.
Therefore, by the classical central limit theorem, we have
\[
n^{-1/2}
\bS_n \to N(\0, \bB)
\]
in distribution, which finishes the proof.
\hfill$\square$

\section{Proof of Theorem 1}

With the preparations in Sections 1 and 2, we now move to the proof of Theorem 1.

Recall that  $\hat\boeta $ satisfies
\begin{equation*}
\frac{\partial H(\hat\boeta)}{\partial \boeta}=\0.
\end{equation*}
Applying the first-order Taylor expansion to $\partial H(\hat\boeta)/\partial\boeta$, and using (\ref{bsn}) and Lemma \ref{lem.A},
we have
\bas
\0
&=&\frac{\partial H(\boeta^*)}{\partial\boeta}+\left(\frac{\partial^2 H(\boeta^*)}{\partial\boeta\partial\boeta^\top} \right) (\hat\boeta-\boeta^*)+o_p(n^{1/2})\\
&=&\bS_n- n\bA (\hat\boeta-\boeta^*)+o_p(n^{1/2}).
\eas
Conditions C1--C4 in the main paper ensure that the matrix $\bA$ is positive definite.
Hence, we obtain an approximation for $\hat\boeta-\boeta^*$ as
\ba
\label{eta.expan}
\hat\boeta-\boeta^*=\left( \begin{array}{c}
\hat\bnu-\bnu^* \\
\hat\rho-\rho^*\\
\hat\btheta-\btheta^*
  \end{array}\right)
=\frac{1}{n} \bA^{-1}\bS_n + o_p(n^{-1/2}).
\ea
This together with the asymptotic property of $\bS_n$ in Lemma \ref{lem.Sn} and Slutsky's theorem gives
\bas
n^{1/2}(\hat\boeta-\boeta^*)
\to N(\0, \bA^{-1}\bB \bA^{-1})
\eas
in distribution,  as $n \to \infty$.

To find the explicit form of $\bA^{-1}\bB \bA^{-1}$, we first identify the structure of $\bA^{-1}$.
Write
\begin{equation*}
    \left(\begin{array}{cc}
       -A_{\rho}  & \bA_{\rho,\bstheta} \\
       \bA_{\bstheta,\rho}  & \bA_{\bstheta}
    \end{array}\right)^{-1} = \left(\begin{array}{cc}
        A^{11} & \bA^{12} \\
        \bA^{21} & \bA^{22}
    \end{array} \right).
\end{equation*}
Using the formula for the inverse of a $2\times2$ block matrix, we have
\bas
A^{11} &=& \{-A_{\rho} - (\bA_{\rho,\bstheta})\bA_{\bstheta}^{-1}(\bA_{\bstheta,\rho})\}^{-1}\\
&=& \left[\{\rho^*(1-\rho^*)\}^{-2}\bfe^\top \bA_{\bstheta}\bfe - \Delta^*\{\rho^*(1-\rho^*)\}^{-1} -  \{\rho^*(1-\rho^*)\}^{-2}\bfe^\top\bA_{\bstheta}\bfe\right]^{-1}\\
&=& -\frac{\rho^*(1-\rho^*)}{\Delta},\\
\bA^{12} &=& (\bA^{21})^\top = -A^{11}(\bA_{\rho,\bstheta})\bA_{\bstheta}^{-1}=\frac{\bfe^\top}{\Delta^*},\\
\bA^{22} &=& \bA_{\bstheta}^{-1} + \bA_{\bstheta}^{-1}(\bA_{\bstheta,\rho})A^{11}(\bA_{\rho,\bstheta})\bA_{\bstheta}^{-1}
= \bA_{\bstheta}^{-1} - \frac{\bfe\bfe^\top}{\Delta^*\rho^*(1-\rho^*)}.
\eas
Hence, $\bA^{-1}$ is given by
\begin{equation}
\label{A_inverse}
    \bA^{-1}
=\left( \begin{array}{cccc}
 \bA_{\bsnu}^{-1} & \0 & \0  \\
 \0 & -\frac{\rho^*(1-\rho^*)}{\Delta^*} & \frac{\bfe^\top}{\Delta^*}  \\
 \0 & \frac{\bfe}{\Delta^*} & \bA_{\bstheta}^{-1} - \frac{\bfe\bfe^\top}{\Delta^{*}\rho^*(1-\rho^*)}
  \end{array}\right).
\end{equation}

With the form of $\bA^{-1}$ in \eqref{A_inverse} and the form of $\bB$ in Lemma \ref{lem.Sn},
after tedious algebra work, we find that
\bas
\bLambda
=
\bA^{-1}\bB \bA^{-1}
=
\left( \begin{array}{ccc}
 \bA_{\bsnu}^{-1} & \rho^*(1-\rho^*) \bA_{\bsnu}^{-1}\bW^\top & \0 \\
 \rho^*(1-\rho^*) \bW\bA_{\bsnu}^{-1} & \rho^*(1-\rho^*) \{\frac{1}{\Delta^*}-S \rho^*(1-\rho^*) \} & \0 \\
 \0 & \0 & \bA_{\bstheta}^{-1}-\frac{\bfe\bfe^\top}{\Delta^{*}\rho^*(1-\rho^*)} \\
 \end{array}\right).
\eas
Recall $S=w^{-1}+(1-w)^{-1}$. Some algebra work leads to
$$
\frac{1}{\Delta^*}-S \rho^*(1-\rho^*)= \frac{1}{\Delta^*}\{\rho^*\nu_0^*+(1-\rho^*)\nu_1^*\}.
$$
This completes the proof of Theorem 1.

\section{Proof of Theorem 2}

Recall that a class of general parameter vector $\bpsi$ of length $p$ is defined as
$$
\bpsi= \int_0^\infty \bu(x;\bnu,\btheta) dG_0(x),
$$
where $\bu(x;\bnu,\btheta)=\left(u_1(x;\bnu,\btheta),\ldots,u_p(x;\bnu,\btheta)\right)^\top$ is a given $p\times 1$ dimensional function.
The MELE of $\bpsi$ is given by
\bas
\hat\bpsi
&=& \Sum \hat p_{ij}\bu(X_{ij};\hat\bnu,\hat\btheta)  \\
&=& \frac{1}{n_{01}+n_{11}}\Sum\frac{\bu(X_{ij};\hat\bnu,\hat\btheta)}{1 + \hat\rho[\exp\{\hat\alpha + \hat\bbeta^{\top} \bq(X_{ij})\}-1] }\\
&=& \frac{1}{nw(1-\hat\nu_0)+n(1-w)(1-\hat\nu_1)}\sum_{ij}\frac{\bu(X_{ij};\hat\bnu,\hat\btheta)}{1 + \hat\rho[\exp\{\hat\alpha + \hat\bbeta^{\top} \bq(X_{ij})\}-1]}I(X_{ij}>0).
\eas

Note that $\hat\bpsi$ is a function of $\hat \boeta$. Hence we write $\hat\bpsi$  as $\hat\bpsi(\hat\boeta)$.
From Theorem 1, we have $\hat\boeta=\boeta^*+O_p(n^{-1/2})$.
Applying the first-order Taylor expansion to $\hat\bpsi(\hat\boeta)$, we get
\bas
\hat\bpsi
&=&
\hat\bpsi(\boeta^*)+ \left(\frac{\partial \hat \bpsi(\boeta^*)}{\partial \boeta}\right)(\hat\boeta-\boeta^*)+o_p(n^{-1/2}).
\eas
For convenience, we write $\bu(x) = \bu(x;\bnu^*,\btheta^*)$.
Note that
\bas
\frac{\partial \hat \bpsi(\boeta^*)}{\partial \boeta}
=
\left(\frac{\partial \hat \bpsi(\boeta^*)}{\partial \bnu}, \frac{\partial \hat \bpsi(\boeta^*)}{\partial \rho}, \frac{\partial \hat \bpsi(\boeta^*)}{\partial \btheta}\right)
\eas
where
\bas
\frac{\partial \hat\bpsi(\boeta^*)}{\partial \bnu}
&=& \frac{1}{n\Delta^{*2}}\sum_{ij}\left\{\frac{\partial \bu(X_{ij};\bnu^*,\btheta^*)/\partial \bnu}{h(X_{ij})}\Delta^* +
(w,1-w) \otimes \frac{\bu(X_{ij})}{h(X_{ij})} \right\}I(X_{ij}>0), \\
\frac{\partial \hat\bpsi(\boeta^*)}{\partial \rho}
&=& -\frac{1}{n\Delta^*}\sum_{ij}\frac{\bu(X_{ij})\{\omega(X_{ij})-1\}}{h(X_{ij})^2}I(X_{ij}>0),\\
\frac{\partial \hat\bpsi(\boeta^*)}{\partial \btheta}
&=& \frac{1}{n\Delta^*}\sum_{ij}\frac{\{\partial \bu(X_{ij};\bnu^*,\btheta^*)/\partial\btheta\} \cdot h(X_{ij})- \bu(X_{ij}) \rho^*\omega(X_{ij})\bQ(X)^\top}{h(X_{ij})^2}I(X_{ij}>0),
\eas
where $\otimes$ refers to Kronecker product.
By the law of large numbers and Lemma \ref{lem2}, we have
\bas
\frac{\partial \hat \bpsi(\boeta^*)}{\partial \boeta}
\to \bC
\eas
in probability, where $\bC=(\bC_{\bsnu},\bC_\rho,\bC_{\bstheta})$ with
\bas
\bC_{\bsnu}
&=& E_0\left\{ \frac{\partial \bu(X;\bnu^*,\btheta^*)}{\partial\bnu} \right\} +  \left(w,1-w\right) \otimes \frac{\bpsi^*}{\Delta^*},\\
\bC_\rho
&=& -E_0\left\{ \frac{\bu(X) \{\omega(X)-1\}}{h(X)}\right\}=\frac{\rho^*\bpsi^* - E_0\left\{h_1(X)\bu(X)\right\}}{\rho^*(1-\rho^*)},\\
\bC_{\bstheta}
&=&
E_0\left[\frac{\{\partial\bu(X;\bnu^*,\btheta^*)/\partial\btheta\} \cdot h(X)- \bu(X) \rho^*\omega(X)\bQ(X)^\top}{h(X)}\right]\\
&=&
E_0\left\{\frac{\partial\bu(X;\bnu^*,\btheta^*)}{\partial\btheta}\right\}- E_0\left\{
h_1(X)\bu(X)\bQ(X)^\top\right\}.
\eas

For convenience, we let
$$
{\bf E}_{0\bsu}= E_0 \{h_0(X)\bu(X) \}\mbox{ and }{\bf E}_{1\bsu}= E_0 \{h_1(X)\bu(X) \}.
$$
Then ${\bf E}_{0\bsu}+{\bf E}_{1\bsu}=\bpsi^*$ and
$$
\bC_\rho= \frac{\rho^*\bpsi^* -  {\bf E}_{1\bsu} }{\rho^*(1-\rho^*)}.
$$

Recall from (\ref{eta.expan}), $\hat\boeta-\boeta^*=n^{-1}\bA^{-1}\bS_n+o_p(n^{-1/2})$.
Therefore,
as $n\to\infty$, $n^{1/2}(\hat\bpsi-\bpsi^*)$ has the same limiting distribution as
\ba
\label{psi.expan}
n^{1/2}\left[\left\{\hat\bpsi(\boeta^*)-\bpsi^* \right\} + \bC\bA^{-1}\bS_n/n\right].
\ea
It can be easily verify that (\ref{psi.expan}) has expectation zero.
In the following, we will decompose its asymptotic variance into  three parts.

Note that the first term of (\ref{psi.expan}) involves
\bas
\hat \bpsi(\boeta^*)
=
\frac{1}{n\Delta^*}\sum_{ij}\frac{\bu(X_{ij})}{h(X_{ij})}I(X_{ij}>0).
\eas
Then the variance of the first term in (\ref{psi.expan}) is
\ba
\nonumber\bGamma_1
&=&  \frac{1}{\Delta^*}E_0\left\{\frac{\bu(X)\bu(X)^\top}{h(X)}\right\} -\frac{1}{w}E_0 \{h_0(X)\bu(X) \}E_0\{h_0(X)\bu(X)^\top\}\\
\nonumber&& -\frac{1}{1-w}E_0\{h_1(X)\bu(X)\}E_0\{h_1(X)\bu(X)^\top\}\\
&=&\frac{1}{\Delta^*}E_0\left\{\frac{\bu(X)\bu(X)^\top}{h(X)}\right\}- \frac{1}{w} {\bf E}_{0\bsu}{\bf E}_{0\bsu}^\top
- \frac{1}{1-w} {\bf E}_{1\bsu}{\bf E}_{1\bsu}^\top,
\label{bgamma1}
\ea
where in the  first step, we have used the results in Lemma \ref{lem2}, and in the second step, we have used the notations ${\bf E}_{0\bsu}$ and ${\bf E}_{1\bsu}$.


Next, we dervie the variance of the second term in (\ref{psi.expan}):
\bas
\bGamma_2
= n\var(\bC\bA^{-1}\bS_n/n)
=\bC \bLambda \bC^\top.
\eas
Together with the form of $\bLambda$ in Theorem 1, we have
\bas
\bGamma_2
&=&
\bC_{\bsnu} \bA_{\bsnu}^{-1}\bC_{\bsnu}^\top + \rho^*(1-\rho^*)\bC_{\bsnu} \bA_{\bsnu}^{-1}\bW^\top \bC_\rho^\top+\rho^*(1-\rho^*)\bC_\rho\bW \bA_{\bsnu}^{-1}\bC_{\bsnu}^\top\\
&& +(\Delta^*)^{-1}\rho^*(1-\rho^*) \{\rho^*\nu_0^*+(1-\rho^*)\nu_1^*\} \bC_\rho \bC_\rho^\top
+ \bC_{\bstheta} \left\{\bA_{\bstheta}^{-1}-\frac{\bfe\bfe^\top}{\Delta^{*}\rho^*(1-\rho^*)}\right\} \bC_{\bstheta}^\top.
\eas
Note that
$$
\bW  \bA_{\bsnu}^{-1} \bW^\top=\frac{\rho^*\nu_0^*+(1-\rho^*)\nu_1^* }{\Delta^*\rho^*(1-\rho^*)}.
$$
Then
\ba\label{bgamma2}
\nonumber
\bGamma_2
&=&
\{ \bC_{\bsnu}+\rho^*(1-\rho^*)\bC_\rho\bW \}  \bA_{\bsnu}^{-1}\{ \bC_{\bsnu}+\rho^*(1-\rho^*)\bC_\rho\bW \} ^\top\\
&&- \frac{1}{\Delta^{*}\rho^*(1-\rho^*)}(\bC_{\bstheta} \bfe) (\bC_{\bstheta} \bfe)^\top +  \bC_{\bstheta}  \bA_{\bstheta}^{-1}  \bC_{\bstheta}^\top.
\ea

Lastly, we derive the covariance of the two terms in  (\ref{psi.expan}). That  is,
\bas
\bGamma_3=n\cov[\bpsi(\boeta^*), n^{-1}\{\bC\bA^{-1}\bS_n\}^\top]
= \cov\{\bpsi(\boeta^*), \bS_n^\top\}\bA^{-1}\bC^\top.
\eas
For convenience,
we denote
$
\cov\{\bpsi(\boeta^*), \bS_n^\top\}=(\bD_{\bsnu}, \bD_{\rho}, \bD_{\bstheta})$.

We first look at
\bas
&&
\cov\{\bpsi(\boeta^*), \bS_{n,\nu_1}\}\\
&=&
\cov\left\{ \frac{1}{n\Delta^*}\sum_{ij}\frac{\bu(X_{ij})}{h(X_{ij})}I(X_{ij}>0),  \frac{-n_{11}}{\nu_1^*(1-\nu_1^*)} \right\}\\
&=&
\frac{-1}{n\Delta^* \nu_1^*(1-\nu_1^*)} \cov\left\{ \sum_{j=1}^{n_1}\frac{\bu(X_{1j})}{h(X_{1j})}I(X_{1j}>0), \sum_{j=1}^{n_1} I(X_{1j}>0) \right\}\\
&=&
\frac{-n_1}{n\Delta^* \nu_1^*(1-\nu_1^*)} \left[ (1-\nu_1^*)E_0\left\{ \frac{\bu(X)\omega(X)}{h(X)} \right\} - (1-\nu_1^*)^2E_0\left\{ \frac{\bu(X)\omega(X)}{h(X)} \right\} \right]\\
&=&
\frac{-(1-w)}{\Delta^*} E_0\left\{  \frac{\bu(X)\omega(X)}{h(X)} \right\}\\
&=&
-(1-\nu_1^*)^{-1} {\bf E}_{1\bsu}
\eas
Similarly, we find
\bas
\cov\{\bpsi(\boeta^*),\bS_{n,\nu_0}\}
=
-(1-\nu_0^*)^{-1} {\bf E}_{0\bsu}.
\eas
Hence
\bas
\label{form.dnu}
\bD_{\bsnu}
=
\left(
-(1-\nu_0^*)^{-1} {\bf E}_{0\bsu}, -(1-\nu_1^*)^{-1}{\bf E}_{1\bsu} \right).
\eas

We can find the expression of $\bD_{\rho}$ and $\bD_{\bstheta}$ in a similar manner.
For $\bD_{\rho}$,
\bas
\bD_{\rho} &=&
\cov\{\bpsi(\boeta^*),S_{n,\rho}\}\\
&=&
-\frac{1}{n\Delta^*} \cov\left\{ \sum_{ij}\frac{\bu(X_{ij})}{h(X_{ij})}I(X_{ij}>0),  \sum_{ij}\frac{\omega(X_{ij})-1}{h(X_{ij})} I(X_{ij}>0) \right\}\\
&=&
C_\rho
+ \frac{\Delta^{*}}{w} E_0\{h_0(X)\bu(X) \}E_0[h_0(X)\{\omega(X)-1\} ]\\
&&
+ \frac{\Delta^{*}}{(1-w)} E_0\{h_1(X)\bu(X) \}E_0[h_1(X)\{\omega(X)-1\}]\\
&=&
C_\rho - \Delta^{*}\bm E_0[h_1(X)\{\omega(X)-1\}],
\eas
where $\bm = \bpsi^*/w-SE_0\{h_1(X)\bu(X)\}=\bpsi^*/w-{\bf E}_{1\bsu}/\{w(1-w)\}$.

For $\bD_{\bstheta}$,
\bas
\bD_{\bstheta} &=&
\cov\{\bpsi(\boeta^*), \bS_{n,\bstheta}^\top\}\\
&=&
\frac{1}{n\Delta^*} \cov\left\{ \sum_{ij}\frac{\bu(X_{ij})}{h(X_{ij})}I(X_{ij}>0),  \sum_{j=1}^{n_1}\bQ(X_{1j})^\top I(X_{1j}>0) - \sum_{ij} h_1(X_{ij})\bQ(X_{ij})^\top I(X_{ij}>0) \right\}\\
&=&
\frac{1}{n\Delta^*} \cov\left\{ \sum_{ij}\frac{\bu(X_{ij})}{h(X_{ij})}I(X_{ij}>0),  \sum_{j=1}^{n_1}h_0(X_{1j})\bQ(X_{1j})^\top I(X_{1j}>0) \right\}\\
&&
- \frac{1}{n\Delta^*} \cov\left\{ \sum_{ij}\frac{\bu(X_{ij})}{h(X_{ij})}I(X_{ij}>0),  \sum_{j=1}^{n_{0}}h_1(X_{0j})\bQ(X_{0j})^\top I(X_{0j}>0) \right\}\\
&=&
(1 - \rho^*) \Delta^{*} \bm E_0\{h_1(X) \bQ(X)^\top\}\\
&=&\bm (\bA_{\bstheta}\bfe)^\top.
\eas

With the form of $ (\bD_{\bsnu}, \bD_{\rho}, \bD_{\bstheta})$ and  the form of $\bA^{-1}$ in \eqref{A_inverse}, $\bGamma_3$ is given as
\bas
\bGamma_3
&=& (\bD_{\bsnu}, \bD_{\rho}, \bD_{\bstheta})\bA^{-1}\bC^\top\\
&=&
\bD_{\bsnu} \bA_{\bsnu}^{-1}\bC_{\bsnu}^\top -\frac{\rho^*(1-\rho^*)}{\Delta^*}\bD_{\rho} \bC_{\rho}^\top + \bD_{\rho}\frac{\bfe^\top}{\Delta^*}\bC_{\bstheta}^\top
 +\bD_{\bstheta}\frac{\bfe}{\Delta^*}\bC_{\rho}^\top + \bD_{\bstheta}\left\{\bA_{\bstheta}^{-1} - \frac{\bfe\bfe^\top}{\Delta^{*}\rho^*(1-\rho^*)}\right\}\bC_{\bstheta}^\top\\
&=& \bD_{\bsnu} \bA_{\bsnu}^{-1}\bC_{\bsnu}^\top  +
\bD_{\bstheta}\bA_{\bstheta}^{-1}\bC_{\bstheta}^\top
 +\frac{1}{\Delta^*}\left\{\bD_{\bstheta}\bfe -\rho^*(1-\rho^*)\bD_{\rho}\right\} \bC_{\rho}^\top
 + \left\{ \frac{\bD_{\rho}}{\Delta^*} - \frac{\bD_{\bstheta}\bfe}{\Delta^{*}\rho^*(1-\rho^*)}\right\}\bfe^\top \bC_{\bstheta}^\top.
 \eas
With the forms of $\bD_{\rho}$ and $\bD_{\bstheta}$, we  have
$$
\bD_{\bstheta}\bA_{\bstheta}^{-1}= \bm \bfe ^\top
\mbox{ and }
\frac{1}{\Delta^*}\left\{\bD_{\bstheta}\bfe -\rho^*(1-\rho^*)\bD_{\rho}\right\}
= \rho^*(1-\rho^*) \bm - \rho^*(1-\rho^*) \bC_{\rho}/ \Delta^*.
$$
Hence
 \ba
 \nonumber
 \bGamma_3
&=& \bD_{\bsnu} \bA_{\bsnu}^{-1}\bC_{\bsnu}^\top  +
\bm\bfe^\top \bC_{\bstheta}^\top
+\left(\bm - \frac{\bC_{\rho}}{\Delta^*}\right)\left\{\rho^*(1-\rho^*)\bC_{\rho}^\top-\bfe^\top \bC_{\bstheta}^\top\right\}\\
\label{bgamma3}&=& \bD_{\bsnu} \bA_{\bsnu}^{-1}\bC_{\bsnu}^\top
+\left(\bm - \frac{\bC_{\rho}}{\Delta^*}\right)\rho^*(1-\rho^*)\bC_{\rho}^\top + \frac{1}{\Delta^*}\bC_{\rho}\bfe^\top \bC_{\bstheta}^\top.
\ea

Putting  $\bGamma_2$ in \eqref{bgamma2} and $\bGamma_3$ in \eqref{bgamma3}  together and using the facts that
$
\bC_{\bstheta}=\mathcal{M}_3
$,
\begin{equation}
\label{m1_relation}
  \bC_{\bsnu}+\rho^*(1-\rho^*)\bC_\rho\bW+ \bD_{\bsnu}=\mathcal{M}_1,
\end{equation}
and
$$
- \frac{(\bC_{\bstheta} \bfe) (\bC_{\bstheta} \bfe)^\top}{\Delta^{*}\rho^*(1-\rho^*)}
+ \frac{1}{\Delta^*}\bC_{\rho}\bfe^\top \bC_{\bstheta}^\top
+ \frac{1}{\Delta^*}\bC_{\bstheta}\bfe \bC_{\rho}^\top
- \frac{\bC_{\rho}}{\Delta^*}\rho^*(1-\rho^*)\bC_{\rho}^\top
=-\frac{\mathcal{M}_2\mathcal{M}_2^\top}{\Delta^* \rho^*(1-\rho^*)},
$$
we have
\ba
\nonumber\bGamma_2 + \bGamma_3+ \bGamma_3^\top&=&( \mathcal{M}_1- \bD_{\bsnu})  \bA_{\bsnu}^{-1}( \mathcal{M}_1- \bD_{\bsnu})^\top-\frac{\mathcal{M}_2\mathcal{M}_2^\top}{\Delta^* \rho^*(1-\rho^*)} +  \mathcal{M}_3 \bA_{\bstheta}^{-1}  \mathcal{M}_3^\top\\
\nonumber&&+  \bD_{\bsnu} \bA_{\bsnu}^{-1}\bC_{\bsnu}^\top +\bC_{\bsnu} \bA_{\bsnu}^{-1} \bD_{\bsnu}^\top
+\rho^*(1-\rho^*)( \bm\bC_{\rho}^\top+\bC_{\rho}\bm^\top)
- \frac{\bC_{\rho}}{\Delta^*} \rho^*(1-\rho^*)\bC_{\rho}^\top\\
\nonumber&=& \mathcal{M}_1   \bA_{\bsnu}^{-1} \mathcal{M}_1^\top-\frac{\mathcal{M}_2\mathcal{M}_2^\top}{\Delta^* \rho^*(1-\rho^*)} +  \mathcal{M}_3 \bA_{\bstheta}^{-1}  \mathcal{M}_3^\top\\
\nonumber&&+  \bD_{\bsnu} \bA_{\bsnu}^{-1}(\bC_{\bsnu}-\mathcal{M}_1)  ^\top +(\bC_{\bsnu}-\mathcal{M}_1) \bA_{\bsnu}^{-1} \bD_{\bsnu}^\top+
 \bD_{\bsnu} \bA_{\bsnu}^{-1}  \bD_{\bsnu}^\top\\
\label{bgamma231}&&
+\rho^*(1-\rho^*)( \bm\bC_{\rho}^\top+\bC_{\rho}\bm^\top)
- \frac{\bC_{\rho}}{\Delta^*} \rho^*(1-\rho^*)\bC_{\rho}^\top.
%
\ea
Next we further simplify the form of $\bGamma_2 + \bGamma_3+ \bGamma_3^\top$.
Note that with (\ref{m1_relation}), we have
\bas
&&\bD_{\bsnu} \bA_{\bsnu}^{-1}(\bC_{\bsnu}-\mathcal{M}_1 )^\top + \bD_{\bsnu} \bA_{\bsnu}^{-1}  \bD_{\bsnu}^\top + \rho^*(1-\rho^*)\bm\bC_{\rho}^\top - \frac{\bC_{\rho}}{\Delta^*} \rho^*(1-\rho^*)\bC_{\rho}^\top\\
&=&-\rho^*(1-\rho^*)\bD_{\bsnu} \bA_{\bsnu}^{-1}\bW^\top\bC_{\rho}^\top  + \rho^*(1-\rho^*)\bm\bC_{\rho}^\top - \frac{\bC_{\rho}}{\Delta^*} \rho^*(1-\rho^*)\bC_{\rho}^\top\\
&=&\rho^*(1-\rho^*)\left(-\bD_{\bsnu} \bA_{\bsnu}^{-1}\bW^\top + \bm -\frac{\bC_{\rho}}{\Delta^*} \right)\bC_{\rho}^\top.
\eas
With the forms of $ \bD_{\bsnu}$, $\bA_{\bsnu}^{-1}$, and $\bW$,
then
\bas
  \bD_{\bsnu}\bA_{\bsnu}^{-1}\bW^\top &=&  -\frac{\nu_0}{\Delta^*(1-\rho^*)}\bpsi^* + \frac{\rho^*\nu_0+(1-\rho^*)\nu_1}{\Delta^*\rho^*(1-\rho^*)}{\bf E}_{1\bsu}\\ &=&-\frac{\nu_0}{\Delta^*(1-\rho^*)}\bpsi^* + \left\{\frac{1}{\Delta^*\rho^*(1-\rho^*)} -S \right\}{\bf E}_{1\bsu}\\
  &=&\frac{1 - \nu_0}{\Delta^*(1-\rho^*)}\bpsi^* - S{\bf E}_{1\bsu} -\frac{1}{\Delta^*\rho^*(1-\rho^*)}\{\rho^*\bpsi^* - {\bf E}_{1\bsu}\}\\
  &=&\bm - \frac{\bC_{\rho}}{\Delta^*}.
\eas
Hence \bas
&&\bD_{\bsnu} \bA_{\bsnu}^{-1}(\bC_{\bsnu}-\mathcal{M}_1 )^\top + \bD_{\bsnu} \bA_{\bsnu}^{-1}  \bD_{\bsnu}^\top + \rho^*(1-\rho^*)\bm\bC_{\rho}^\top - \frac{\bC_{\rho}}{\Delta^*} \rho^*(1-\rho^*)\bC_{\rho}^\top=\0
\eas
and $\bGamma_2 + \bGamma_3+ \bGamma_3^\top$ in \eqref{bgamma231} becomes
\ba
\nonumber
\bGamma_2 + \bGamma_3+ \bGamma_3^\top&=&\mathcal{M}_1   \bA_{\bsnu}^{-1} \mathcal{M}_1^\top-\frac{\mathcal{M}_2\mathcal{M}_2^\top}{\Delta^* \rho^*(1-\rho^*)} +  \mathcal{M}_3 \bA_{\bstheta}^{-1}  \mathcal{M}_3^\top\\
\label{bgamma232}&&+ (\bC_{\bsnu}-\mathcal{M}_1) \bA_{\bsnu}^{-1}\bD_{\bsnu}^\top
+\rho^*(1-\rho^*)\bC_{\rho}\bm^\top.
\ea

With $\bGamma_1$ in \eqref{bgamma1} and $\bGamma_2 + \bGamma_3+ \bGamma_3^\top$ in \eqref{bgamma232},
to show that $\bGamma=\bGamma_1+\bGamma_2 + \bGamma_3+ \bGamma_3^\top $,
we need to argue that
\ba
\label{fpart}
-\frac{\bpsi^*\bpsi^{*\top} }{\Delta^*}
&=&  (\bC_{\bsnu}-\mathcal{M}_1) \bA_{\bsnu}^{-1}\bD_{\bsnu}^\top
+\rho^*(1-\rho^*)\bC_{\rho}\bm^\top
-\frac{1}{w} {\bf E}_{0\bsu}{\bf E}_{0\bsu}^\top
- \frac{1}{1-w} {\bf E}_{1\bsu}{\bf E}_{1\bsu}^\top.
\ea

Note that
$$
\bC_{\bsnu}
= \mathcal{M}_1+  \left(w,1-w\right) \otimes \frac{\bpsi^*}{\Delta^*}.
$$
Then
\ba
\label{fpart1}
(\bC_{\bsnu}-\mathcal{M}_1) \bA_{\bsnu}^{-1}\bD_{\bsnu}^\top&=& -\frac{\nu_0}{\Delta^*}\bpsi{\bf E}_{0\bsu}^\top  -\frac{\nu_1}{\Delta^*}\bpsi{\bf E}_{1\bsu}^\top
=-\bpsi \left(\frac{\nu_0}{\Delta^*} {\bf E}_{0\bsu}^\top + \frac{\nu_1}{\Delta^*} {\bf E}_{1\bsu}\right)^\top.
\ea
Recall that
$$
\rho^*(1-\rho^*)\bC_{\rho}=\rho^*\bpsi^*- {\bf E}_{1\bsu}=\rho^* {\bf E}_{0\bsu}- (1-\rho^*){\bf E}_{1\bsu}
$$
and
$$
\bm =\bpsi^*/w-{\bf E}_{1\bsu}/\{w(1-w)\}={\bf E}_{0\bsu}/w- {\bf E}_{1\bsu}/(1-w).
$$

Then
\ba
\nonumber&&
\rho^*(1-\rho^*)\bC_{\rho}\bm^\top
-\frac{1}{w} {\bf E}_{0\bsu}{\bf E}_{0\bsu}^\top
- \frac{1}{1-w} {\bf E}_{1\bsu}{\bf E}_{1\bsu}^\top\\
\nonumber&=& \left\{\rho^* {\bf E}_{0\bsu}- (1-\rho^*){\bf E}_{1\bsu}\right\} \left\{{\bf E}_{0\bsu}/w- {\bf E}_{1\bsu}/(1-w)\right\}^\top-\frac{1}{w} {\bf E}_{0\bsu}{\bf E}_{0\bsu}^\top
- \frac{1}{1-w} {\bf E}_{1\bsu}{\bf E}_{1\bsu}^\top\\
\nonumber&=&-\frac{1-\rho^*}{w} {\bf E}_{0\bsu}{\bf E}_{0\bsu}^\top
-\frac{1-\rho^*}{w}{\bf E}_{1\bsu}{\bf E}_{0\bsu}^\top
-\frac{\rho^*}{1-w} {\bf E}_{0\bsu}{\bf E}_{1\bsu}^\top
-\frac{\rho^*}{1-w} {\bf E}_{1\bsu}{\bf E}_{1\bsu}^\top\\
\nonumber&=&-({\bf E}_{0\bsu}+{\bf E}_{1\bsu} ) \left(\frac{1-\rho^*}{w} {\bf E}_{0\bsu}+\frac{\rho^*}{1-w} {\bf E}_{1\bsu} \right ) ^\top\\
\label{fpart2}&=&-\bpsi^*\left(\frac{1-\nu_0}{\Delta^*} {\bf E}_{0\bsu}+\frac{1-\nu_1}{\Delta^*}  {\bf E}_{1\bsu} \right ) ^\top
\ea
With the fact ${\bf E}_{0\bsu}+{\bf E}_{1\bsu}=\bpsi$, combining \eqref{fpart1} and \eqref{fpart2} gives
\bas
(\bC_{\bsnu}-\mathcal{M}_1) \bA_{\bsnu}^{-1}\bD_{\bsnu}^\top
+\rho^*(1-\rho^*)\bC_{\rho}\bm^\top-\frac{1}{w} {\bf E}_{0\bsu}{\bf E}_{0\bsu}^\top
- \frac{1}{1-w} {\bf E}_{1\bsu}{\bf E}_{1\bsu}^\top = -\frac{1}{\Delta}\bpsi\bpsi^{*\top},
\eas
which verifies \eqref{fpart}.
Hence
$$
\bGamma=\bGamma_1+\bGamma_2 + \bGamma_3+ \bGamma_3^\top.
$$

Applying Slutsky's theorem and the central limit theorem to  (\ref{psi.expan}), we get
$$
n^{1/2}\left( \hat\bpsi-\bpsi^*\right)
\to
N \left(\0, \bGamma \right)
$$
in distribution.
This completes the  proof of Theorem 2.

\section{Proof of Corollary 1}
Note that for $\bpsi=\left(
\bpsi_0^\top,
\bpsi_1^\top
\right)^\top$,
$ \bu(x;\bnu,\btheta)$ can be written as
\begin{equation}
\label{ufun}
    \bu(X;\bnu,\btheta) = \left(\begin{array}{c}
(1-\nu_0) {\bf a}(x)  \\
(1-\nu_1) {\bf a}(x)\omega(x;\btheta)
    \end{array} \right).
\end{equation}
We plug this $ \bu(x;\bnu,\btheta)$
to $\bGamma$ in Theorem 2 to obtain $\bGamma_{semi}$.

Note that
\begin{equation*}
E_0\left\{\frac{\bu(X;\bnu^*,\btheta^*)}{\partial \bnu}\right\} = \left(\begin{array}{cc}
        -\frac{\bpsi_0}{1 - \nu_0^*}&\0 \\
       \0&-\frac{\bpsi_1}{1 - \nu_1^*}
    \end{array} \right)
 \end{equation*}
 and
 \begin{equation*}
    E_0\left\{\frac{\partial\bu(X;\bnu^*,\btheta^*)}{\partial \btheta}\right\} = \left(\begin{array}{c}
        \0 \\
       (1 - \nu_1^*)E_0\{{\bf a}(X)\omega(X)\bQ(X)^\top\}
    \end{array} \right).
\end{equation*}
Then we have $\mathcal{M}_1 = \diag\left\{-\bpsi_0/(1 - \nu_0^*), -\bpsi_1/(1 - \nu_1^*)\right\}$ and
\begin{equation*}
 \mathcal{M}_2 =\left(\begin{array}{c}
       -\rho^*\bpsi_0 \\(1 - \rho^*)\bpsi_1
    \end{array} \right),~~~ \mathcal{M}_3 =\left(\begin{array}{c}
        -w^{-1}\Delta^*(1-\rho^*)E_0\{h_1(X){\bf a}(X)\bQ(X)^\top\} \\(1-w)^{-1}\Delta^*(1-\rho^*)E_0\{h_1(X){\bf a}(X)\bQ(X)^\top\}
    \end{array} \right).
\end{equation*}
Plugging $\mathcal{M}_1$, $\mathcal{M}_2$, and $\mathcal{M}_3$  to $\bGamma$,
and after some simplifications, $\bGamma_{semi}$ is found to be
\begin{eqnarray*}
\bGamma_{semi} &=& \frac{1}{\Delta^*}E_0\left\{\frac{\bu(X)\bu(X)^\top}{h(X)} \right\} -
  \left(\begin{array}{cc}
       \frac{\bpsi_0\bpsi_0^\top}{w} &\0 \\
    \0 & \frac{\bpsi_1\bpsi_1^\top}{1-w}
    \end{array}\right)+    \left(\begin{array}{cc}
        \frac{1}{w^2} & -\frac{1}{w(1-w)} \\
       -\frac{1}{w(1-w)}  & \frac{1}{(1-w)^2}
    \end{array}\right) \otimes \bD_1,
\end{eqnarray*}
where  $$
\bD_1= \{\Delta^*(1-\rho^*)\}^2E_0\{h_1(X){\bf a}(X)\bQ(X)^\top\}\bA_{\bstheta}^{-1}E_0\{h_1(X)\bQ(X){\bf a}(X)^\top\}.
$$

Plugging (\ref{ufun}) to the first term of $\bGamma_{semi}$,
we find that
\bas
\frac{1}{\Delta^*}E_0\left\{\frac{\bu(X)\bu(X)^\top}{h(X)} \right\}
&=&\left(
  \begin{array}{cc}
  w^{-1}( {\bf V}_0+ \bpsi_0\bpsi_0^\top )
 & \0 \\
         \0 & (1-w)^{-1} (  {\bf V}_1 +\bpsi_1\bpsi_1^\top)
    \end{array}
  \right)\\
  &&-
   \left(\begin{array}{cc}
        \frac{1}{w^2} & -\frac{1}{w(1-w)} \\
       -\frac{1}{w(1-w)}  & \frac{1}{(1-w)^2}
    \end{array}\right) \otimes \bD_0,
\eas
where
$$
\bD_0= \Delta^*(1-\rho^*) E_0\{h_1(X){\bf a}(X){\bf a}(X)^\top\}.
$$

Hence
\begin{equation*}
 \bGamma_{semi} =     \bGamma_{non} -
    \left(\begin{array}{cc}
        \frac{1}{w^2} & -\frac{1}{w(1-w)} \\
       -\frac{1}{w(1-w)}  & \frac{1}{(1-w)^2}
    \end{array}\right) \otimes (\bD_0-\bD_1).
\end{equation*}
Recall that  $$
{\bf d}(X)={\bf a}(X)- \Delta^*(1-\rho^*) E_0\left\{h_1(X){\bf a}(X)\bQ(X)^\top\right\} \bA_{\bstheta}^{-1} \bQ(X)
$$
and
$$
\bA_{\bstheta} =
\Delta^*(1-\rho^*)E_0\left[h_1(X) \bQ(X)\bQ^\top(X) \right].
$$
It can be verified that
$$
E_0\left\{
h_1(X) {\bf d}(X){\bf d}(X)^\top
\right\}
=\frac{1}{\Delta^*(1-\rho^*) } (\bD_0-\bD_1 ).
$$
Therefore
 $$
\bGamma_{semi}=
  \bGamma_{non}-\Delta^*(1-\rho^*) E_0\left\{h_1(X)
   \left(
  \begin{array}{c}
  w^{-1} {\bf d}(X)\\
  -(1-w)^{-1}{\bf d}(X)\\
  \end{array}
  \right)
  \left(
    \begin{array}{c}
  w^{-1} {\bf d}(X)\\
  -(1-w)^{-1}{\bf d}(X)\\
  \end{array}
  \right)^\top
  \right\},
  $$
  as claimed in Corollary 1.
This finishes the proof.

\end{document}